\documentclass[12pt]{iopart}
\usepackage{graphicx}
\usepackage{iopams}
\usepackage[squaren]{SIunits} 
\usepackage{harvard}
\usepackage{textcomp}
\usepackage{multirow}
\usepackage{epstopdf}
\usepackage{color}
\usepackage[normalem]{ulem}  
\usepackage{dcolumn} 
\newcolumntype{.}{D{.}{.}{-1}}

\graphicspath{{./figures/}}

\newcommand{\apj}{ApJ}
\newcommand{\apjs}{ApJS}
\newcommand{\apjl}{ApJL}
\newcommand{\mnras}{MNRAS}
\newcommand{\prd}{PhRD}
\newcommand{\prc}{PhRC}
\newcommand{\prl}{PRL}
\newcommand{\npa}{Nucl Phys A}
\newcommand{\nat}{Nature}
\newcommand{\aap}{A\&A}

\newcommand{\adndt}{At. Data Nucl. Data Tables}

\newcommand{\ijmpa}{Int. J. Mod. Phys. A}

\newcommand{\Lnumax}{L_{\nu,{\rm peak}}}
\newcommand{\tmax}{t_{\rm peak}}
\newcommand{\ttouch}{t_{\rm m}}
\newcommand{\tinit}{t_{\rm init}}
\newcommand{\dtcool}{\Delta t_{\rm cool}}
\newcommand{\tendtr}{t_{\rm o}}
\newcommand{\Enumin}{\langle E_{\nu,{\rm min}} \rangle}
\newcommand{\Enumax}{\langle E_{\nu,{\rm max}} \rangle}
\newcommand{\el}{{\rm e}}
\newcommand{\Lnue}{L_{\nu_\el}}
\newcommand{\Lanue}{L_{\bar{\nu}_\el}}
\newcommand{\nue}{\nu_\el}
\newcommand{\anue}{\bar{\nu}_\el}
\newcommand{\Ye}{Y_\el}
\newcommand{\Yp}{Y_{\rm p}}
\newcommand{\Yn}{Y_{\rm n}}
\newcommand{\n}{{\rm n}}
\newcommand{\p}{{\rm p}}
\newcommand{\me}{m_\el}
\newcommand{\kB}{k_{\rm B}}
\newcommand{\F}{\mathcal F}
\newcommand{\rhofreezeout}{\rho_{\rm free}}

\definecolor{APC}{RGB}{0,90,169}
\definecolor{DMC}{RGB}{0,157,129}
\definecolor{AAS}{RGB}{128,0,128}

\begin{document}

\title[The role of weak interactions in dynamic ejecta from BNS mergers]{The role of weak interactions in dynamic ejecta from binary neutron star mergers.}

\author{D. Martin$^{1,2}$, A. Perego$^{3,1,2,4,5}$, W. Kastaun$^{6,7}$, A. Arcones$^{1,2}$}

\address{{~}$^1$ Institut f\"ur Kernphysik, Technische Universit\"at Darmstadt, Schlossgartenstra{\ss}e 2, 64289 Darmstadt, Germany. \\
         {~}$^2$ GSI Helmholtzzentrum f\"ur Schwerionenforschung GmbH, Planckstra{\ss}e 1, 64291 Darmstadt, Germany.}
\address{{~}$^3$ Istituto Nazionale di Fisica Nucleare, Sezione di Milano-Bicocca, Gruppo Collegato di Parma, Parco Area delle Scienze 7/A 43124 Parma, Italy. \\
         {~}$^4$ Universit\`a degli Studi di Milano-Bicocca, Dipartimento di Fisica, Piazza della Scienza 3, 20126 Milano, Italy. \\
         {~}$^5$ Universit\`a degli Studi di Parma, Dipartimento di Matematica, Fisica e Scienze Informatiche, Parco Area delle Scienze 7/A 43124 Parma, Italy.
         }
\address{{~}$^6$ Max Planck Institute for Gravitational Physics (Albert Einstein Institute), Callinstra{\ss}e 38, 30167 Hannover, Germany. \\
         {~}$^7$ Leibniz Universit\"at Hannover, Institute for Gravitational Physics, Callinstra{\ss}e 38, 30167 Hannover, Germany.}
 
\ead{a.perego2@pr.infn.it}
\vspace{10pt}
\begin{indented}
\item[]September 2017
\end{indented}

\begin{abstract}
Weak reactions are critical for the neutron richness of the matter dynamically ejected after the merger of two neutron stars. 
The neutron richness, defined by the electron fraction ($\Ye$), determines which heavy elements are produced by the r-process 
and thus directly impacts the kilonova light curve. In this work, we have performed a systematic and detailed 
post-processing study of the impact of weak reactions on the distribution of the electron fraction and of the entropy 
on the dynamic ejecta obtained from an equal mass neutron star binary merger simulated in full general relativity and with 
microscopic equation of state. Previous investigations indicated that shocks increase $\Ye$, however our results 
show that shocks can also decrease $\Ye$, depending on their thermodynamical conditions. Moreover, we have found that 
neutrino absorption are key and need to be considered in future simulations. 
We also demonstrated that the angular dependence of the neutrino 
luminosity and the spatial distribution of the ejecta can lead to significant difference in the electron fraction distribution. 
In addition to the detailed study of the $\Ye$ evolution and its dependences, we have performed nucleosynthesis calculations.
They clearly point to the necessity of improving the neutrino treatment in current simulations to be able to predict the 
contribution of neutron star mergers to the chemical history of the universe and to reliable calculate their kilonova light curves.
\end{abstract}

%
\vspace{2pc}

\section{Introduction}
\label{sect:intro}

The merger of two neutron stars is intrinsically a multi-messenger event.
The energy that is released by these events produces a large variety
of different transients, all of which are nowadays potentially detectable from the Earth
(see, e.g., \citeasnoun{Rosswog:2015}, \citeasnoun{Baiotti.etal:2017} and \citeasnoun{Fernandez.Metzger:2016} for recent reviews). 
In the case of a sufficiently close event, we expect to observe gravitational waves 
from terrestrial detectors; electromagnetic radiation, 
ranging from gamma- and X-rays related with short gamma-ray bursts, 
to infrared emission associated with a kilonova \cite{Metzger.etal:2010b} (also called 
macronova, \citeasnoun{Kulkarni:2005}); and possibly neutrinos.
Moreover, neutron star mergers play a critical role in
the chemical evolution of the universe as they are most likely
production site of half of the heavy elements via the rapid neutron
capture process \citeaffixed{Lattimer.etal:1977,Eichler.etal:1989,Bauswein.etal:2014}{r-process,}. 
Indeed, the very neutron-rich isotopes
that are formed and ejected in these events power the kilonova light
curve as they decay to stability.

Our understanding of neutron star mergers is increasing very fast as
computer power and observations improve. In current simulations, more
microphysics has been included, namely high density equations of state
\citeaffixed{Hotokezaka.etal:2011,Bauswein.etal:2013,Read.etal:2013,Rezzolla.Takami:2016,Bernuzzi.etal:2016,Radice.etal:2017,Bovard.etal:2017}{e.g.,} 
and neutrinos 
\citeaffixed{Neilsen.etal:2014,Foucart.etal:2015,Sekiguchi.etal:2015,Radice.etal:2016}{e.g.,}.
The transport of neutrinos produced in the 
hot and dense remnant is not full consistently included in current 3D, full general relativistic (GR) 
simulations. However, it has been shown that neutrinos and electron/positron 
captures are crucial to understand the evolution of the neutron richness 
(i.e., of the electron fraction, $\Ye$) of the ejecta. This is the focus 
of our paper. Several questions concerning the evolution of the electron 
fraction in the dynamic ejecta still remain open and largely unexplored: What is the most
relevant process responsible for the change in the electron fraction
of the ejecta? What is the impact of shocks on the electron fraction?
How robust is the r-process nucleosynthesis from binary neutron star (BNS) mergers
under variations of the electron fraction?

Recent GR simulations, including microphysical equation of state (EOS) and neutrino
treatment, indicate that the electron fraction of the dynamic ejecta
can be significantly changed with respect to the initial cold weak
equilibrium values \cite{Wanajo.etal:2014,Sekiguchi.etal:2015,Foucart.etal:2015,Goriely.etal:2015,Radice.etal:2016,Bovard.etal:2017}.
Matter ejected by tidal interaction is expected to stay
relatively cold. Therefore, in the absence of strong neutrino
irradiation, its electron fraction should not change significantly \citeaffixed{Korobkin.etal:2012}{e.g.,}. 
In contrast, matter ejected by shocks (occurring when the two neutron
stars collide or after the remnant has formed) is heated up to
significantly large temperatures.  Under these conditions,
electron-positron pairs are copiously produced and weak processes
involving neutrinos can alter the initial electron fraction. In
particular, positron captures on free neutrons can increase the
electron fraction. Similarly, neutrino irradiation can enhance the
electron fraction through the absorption of electron neutrinos on the
initially very neutron-rich ejecta.

In this paper, we consider trajectories of shock heated ejecta
obtained from a GR hydrodynamical simulation \cite{Kastaun.etal:2016b}.
We include the impact of neutrino emission and absorption in a post-processing step, similar
to \citeasnoun{Goriely.etal:2015}, but including also consistently the consequences of 
weak reactions on the entropy evolution. 
Our approach allows to explore the role of individual 
weak reactions on the final distributions of electron fraction and 
entropy, and, in turn, on the detailed nucleosynthesis abundances.  
Our results indicate that the occurrence of a shock in the ejection process does
{\em not necessary} lead to an increase of the electron fraction in
neutron-rich matter, as found in previous works for other types of shock heated ejecta 
\citeaffixed{Sekiguchi.etal:2015}{e.g.,}.
If this is the case, neutrino absorption plays a major role in increasing
$\Ye$. We also study the dependence on the intensity of the neutrino irradiation and found a
non-negligible effect on the nucleosynthesis. For the first time, we
present the impact of neutrino emission anisotropies that can become
very important and have strong consequences on the prediction of the
final abundances and thus kilonova light curve.

The paper is structured as follows: In Sec.~\ref{sect:simulation}, we summarize the properties of the GR simulation that provides 
the dynamical evolution of the BNS merger, and we present the properties of the the ejecta obtained in that simulation 
and of the neutrino emission that we computed from the merger profiles. 
In Sec.~\ref{sect:weak-reactions}, we introduce the post-processing treatment for the weak reactions to evolve
the electron fraction and entropy of each tracer. Sec.~\ref{sect:results} is devoted to the presentation and discussion 
of our results in terms of distributions
of the properties of the ejecta and of abundances of the nucleosynthesis yields. Finally, in Sec.~\ref{sect:conclusion}, we draw our conclusions.

\section{Binary NS merger simulation}
\label{sect:simulation}
\subsection{Setup and numerical evolution} \label{sec:setup}
The ejected matter data studied in this work are extracted from one of the numerical simulations
described in \citeasnoun{Kastaun.etal:2016b}. The \texttt{SHT\_UU} model consists of two NSs with 
gravitational mass of $1.4\,M_\odot$ each, and employs the EOS by Shen, Horowitz, Teige 
\citeaffixed{Shen.etal:2010,Shen.etal:2011}{SHT, }. The initial NS spins are aligned with the orbital angular 
momentum, with dimensionless spin $J/M^2 \approx 0.125$. We note that the other spin
configurations studied in \citeasnoun{Kastaun.etal:2016b} did not yield a significant amount of ejecta.
For the SHT EOS, the maximum possible baryonic (gravitational) mass of a non-rotating NS
(at zero temperature in $\beta$-equilibrium) is $3.33\usk M_\odot$ ($2.77\usk M_\odot$), which is unusually large.
Since the total baryonic mass of the system is below this critical value, the merger remnant
is a stable NS (which we regard as an astrophysical corner case).

The numerical evolution was carried out in general relativity and the matter was treated
as a perfect fluid. Pressure, internal energy, and specific entropy were computed from density, 
temperature, and electron fraction using a three-dimensional interpolation table of the SHT EOS,
as described in \citeasnoun{Galeazzi.etal:2013}. We did not include magnetic fields (which might 
drive further matter outflows in addition to dynamical ejecta, \citeasnoun{Siegel.etal:2014}).
The code made use of moving box mesh refinement, with a finest resolution of
$295 \usk\meter$, and the outer boundary is located at $945\usk\kilo\meter$.
The numerical methods and code set-up are described in detail in \citeasnoun{Kastaun.etal:2016b}. 
Neutrino radiation was not taken into account in the
simulation itself, and the electron fraction was passively advected with the fluid.
In this study, we account for the neutrino physics in a post-processing
step, which will be described in Sec.~\ref{sec:nu_properties}, neglecting any impact on the 
fluid dynamics in the remnant.

One technical detail relevant for this work is the usage of an artificial atmosphere, which is
a standard method where a minimum mass density is enforced throughout the computational domain,
and the velocity is set to zero inside the artificial atmosphere. This
is required because the hydrodynamic equations degenerate in vacuum and also because our 
tabulated EOS does not extend to arbitrary low values. Due to the latter, we used a 
relatively dense atmosphere of $6\times 10^7~{\rm g \, cm^{-3}}
$ (with temperature 
$0.06 \usk\mega\electronvolt$ and electron fraction 0.4).
Although such an atmosphere only weakly affects the dynamics of the inspiral, merger remnant, and 
disk, it has a strong impact on the low-density ejecta 
\citeaffixed{Endrizzi.etal:2016,Kastaun.etal:2015}{see the discussion in}. 
We stress that, since the goal of this work is to study the impact of weak reactions on the
properties of the ejecta, the potential effect of the artificial atmosphere on the total amount 
of ejected material and (more importantly) the escape velocity, has no relevance for us. 
That said, we do make use of a method developed to correct for the drag, which will be described 
in \ref{sec:drag}.

\subsection{Ejecta and tracers} 
\label{sec:tracer}

Extracting the trajectories of ejected matter from a numerical simulation is a nontrivial task 
that requires various approximations. In the following, we describe the main difficulties and the 
solutions used for this work. 
The first challenge is that numerical simulations can only be run on short time scales, 
necessitating a criterion to judge if a given fluid element will reach infinity eventually.
For this, we use the standard approach of assuming geodesic motion, approximating the spacetime 
as stationary. This results in the condition $u_t<-1$, where $u$ is the fluid 4-velocity 
\citeaffixed{Kastaun.Galeazzi:2015}{e.g.,}.
Although physically we do not expect significant deviations from Keplerian motion once the 
ejected matter is expelled from the vicinity of merger remnant and disk, the artificial 
atmosphere (see previous section) causes an unphysical drag force. In our case, 
the impact is quite strong because of the low ejecta mass (and hence density) and the 
high atmosphere density. Almost all matter that was
unbound at some point becomes bound again at large radii. 
We attribute this to the spurious drag force
based on an animation (available in the supplemental material of \citeasnoun{Kastaun.etal:2016b})
showing how regions of unbound matter run into the artificial atmosphere 
and slowly dissolve. In \citeasnoun{Kastaun.etal:2016b}, we measured the flux 
of unbound matter through spherical surfaces of increasing radius and used the maximum
as an estimate for the amount of ejected matter. For this work however, we require
the trajectories of ejected matter up to radii much larger than the simulation domain.
Therefore, we extrapolated trajectories assuming Keplerian motion when
the drag force becomes relevant. To determine this correction, we constructed a simple model for the 
atmospheric drag, which we also use to validate that the artificial atmosphere is the reason
for the decrease of unbound mass at large radius. The details of this model are presented in~\ref{sec:drag}.

The next challenge is to extract fluid trajectories from the numerical simulation, which
uses a numerical grid to describe the fluid (in contrast to smoothed particle hydrodynamics 
codes, for which trajectories are an integral part of the evolution). One approach 
would be to follow tracers during the evolution \cite{Bovard.Rezzolla:2017}. 
A difficulty with that method is the placement
of the initial tracer positions. Since only a small fraction of the fluid mass ends up as 
ejected matter, a large fraction of tracers will be wasted. Achieving a good coverage of ejected
matter becomes computationally challenging. To overcome this limitation, we extracted the fluid
trajectories in a post-processing step from 3D data saved during the evolution.
One shortcoming of this procedure is that the data need to be stored with sufficiently high 
resolution in both space and time to maintain accuracy in the integration.
However, it has the advantage that we can identify unbound material at a suitable time 
and then trace its movement both forward and backward in time. In our case, most matter 
became unbound during a few $\milli\second$. This allowed us to simplify the code and start 
the time integration 
of all tracers at the same seed time, where the amount of unbound matter becomes maximal. 
The seed positions are taken from a regular grid, and we assign the mass proportionally to the local density. 
Only grid points with $u_t < -1 + \delta$ are used as seed
positions. We lower the threshold for the unbound matter criterion by some suitable $\delta>0$ 
to make up for the fact that some matter becomes unbound later or re-bound earlier.
After computing the trajectories we remove the ones that never became unbound.
Starting from the seed positions, we integrate both forward and backward in time,
using a second order scheme and cylindrical coordinates. A further complication of the artificial
atmosphere is that the least dense parts of the ejecta fall below the cutoff density
at some point, and become part of the non-moving atmosphere. We only keep trajectories 
that can be traced to the end.
Finally, we combine neighboring trajectories in order to limit 
the number of trajectories to ${\approx}1000$, since the nuclear network calculations using 
the trajectories are more expensive than the extraction. We have tested that this procedure has
practically no impact on our analysis.

\begin{figure}
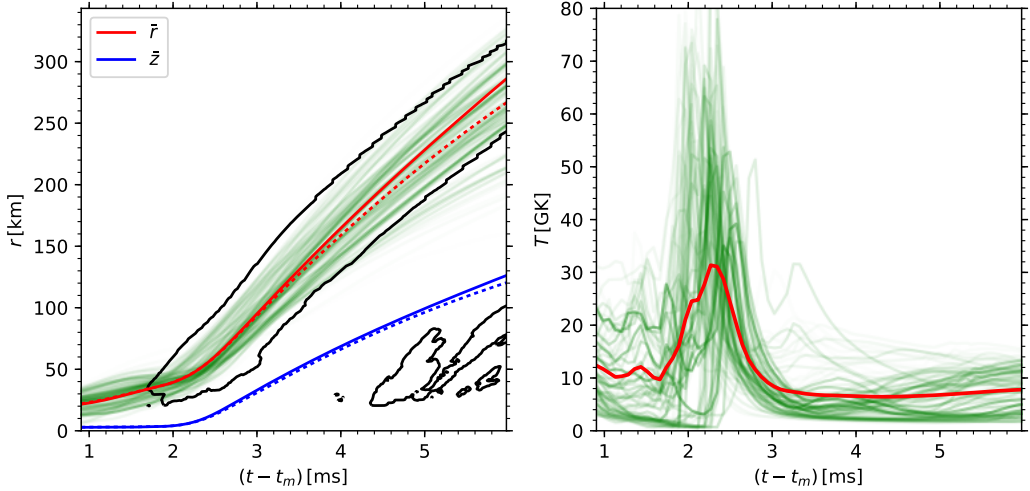

  \begin{center}
    \includegraphics[width=0.9\columnwidth]{{{kick_alt}}}  
    \caption{Properties of the ejected tracers versus time after merger, $\ttouch$. 
    Left: spherical radial coordinate and $z$ coordinate. The solid blue and red lines 
    represent the 
    mass-weighted average of the tracer positions after correcting for the atmospheric drag,
    the dotted lines the same without the correction. The transparent lines show individual 
    tracer trajectories (opacity proportional to mass). The contour line shows regions with 
    unbound material (no drag correction) obtained directly from the numerical grid by 
    summing unbound mass into histograms over radial bins at regular times.
    Right: Matter temperature.
    Thick black lines represent averages over the whole tracer sample (weighting each 
    individual trajectory by its mass). The transparent lines represent individual trajectories.}
    \label{fig:kick}
  \end{center}
\end{figure}

We find that the ejecta for the model at hand are neither tidally ejected nor part of 
a breakout shock formed when the stars merge. Instead, the ejecta originate from the inner 
part of the disk. Around $2\usk\milli\second$ after merger, one of the remnant oscillations 
sends a wave into the disk, which liberates the marginally bound
matter. The ejecta form two concentric rings above and below the orbital plane, which expand 
radially (and also in $z$-direction).
We note that it does not require any violent movement of the remnant to eject matter, because 
the density scales are very different. In our example, it appears that a wave steepens 
into a shock. At least, we found a steep increase of specific entropy (from a few 
to ${\sim} 7 \, \kB \, \mathrm{baryon}^{-1}$) at the same time the ring starts accelerating 
outwards.
The average ejecta temperature (entropy) is increased from about 10\,GK to about 30\,GK
by the shock heating, and then cools down adiabatically while the density is decreasing.
It is important to mention that most material is still classified as bound when the 
temperature has already dropped below NSE conditions. Since the temperature enters 
into the initial conditions for nucleosynthesis calculations, it is necessary to know
the thermal history of unbound matter. Using the conditions found at the time where 
matter becomes unbound is only sufficient for tidally ejected (i.e., cold) matter.
Mass-weighted average temperature and radius for the (Kepler-extrapolated) tracers 
are shown in Figure~\ref{fig:kick}. The correlation between acceleration and temperature
increase is clearly visible. Note that near the end the temperature slightly increases 
again. We assume that in addition to the drag, the interaction with the artificial 
atmosphere also causes heating. 
However, before performing our analysis we correct this artificial increase, by assuming constant entropy
outside a fiducial radius of 200 km.
Finally, as an independent validation of the correct tracing of fluid elements, we compute
the radial extent of unbound matter at each time directly from data on the numerical grid.
For this, we collect at regular time intervals the unbound mass in histograms
with bins corresponding to the radial coordinate. The resulting unbound regions
are also shown in Figure~\ref{fig:kick}. The bulk of the trajectories are clearly 
inside this region, although, as described above, some trajectories also become 
bound again. The Figure also shows the impact of the drag correction, which will be 
described in \ref{sec:drag}. 

\subsection{Neutrino properties}
\label{sec:nu_properties}

Numerical simulations of BNS mergers, including neutrino emission, show characteristic trends in the evolution
of the neutrino luminosities, $L_{\nu}$, and mean energies, $\langle E_{\nu} \rangle$ 
\cite{Ruffert.etal:1997,Rosswog.Liebendoerfer:2003,Neilsen.etal:2014,Foucart.etal:2015,Sekiguchi.etal:2015,Radice.etal:2016}. 
While the emission of radiation is negligible during the cold
inspiral phase, as soon as the two NSs touch ($t=\ttouch$) and the temperature in the remnant increases, 
neutrino luminosities are boosted and reach their peak values (a few $10^{53} \, {\rm erg \, s^{-1}}$) 
within a few ms ($t= \tmax$). Additionally, the neutrino mean energies rise, as a consequence of the temperature growth.
After a few oscillations (lasting typically no more than a few ms, and related with the motion of the merging 
cores and the formation of the disk), both the luminosities and the mean energies settle to almost stationary values. 
On a longer time scale, the luminosities slowly decrease as a result of the remnant and disk cooling, and of the reduction of 
the accretion rate inside the disk. To mimic this behavior, we model the neutrino luminosities according 
to the following analytic prescription:
\begin{equation}
L_{\nu}(t) =
 \left\{
 \begin{array}{ll}
  0 & {\rm for} \quad t \leq \ttouch, \\
  \Lnumax \, \left( \frac{t - \ttouch}{\tmax-\ttouch} \right) & {\rm for } \quad  \ttouch < t \leq \tmax,  \\
  \Lnumax \, \exp{\left( - \frac{t-\ttouch}{\dtcool} \right)} & {\rm for } \quad t > \tmax, 
 \end{array}
 \right.
 \label{eq: analytic_lum_evolution}
\end{equation}
where we have included the growth of the luminosities between $\ttouch$ and $\tmax$, and the subsequent decrease, but have 
neglected the transient oscillations. 
In the previous equation, $t$ is the time with respect to the beginning of the simulation. 
We consider a typical cooling time scale $\dtcool \sim 500 \,{\rm ms}$, 
comparable with the disk life time and with the diffusion time scale from the central MNS \citeaffixed{Dessart.etal:2009,Perego.etal:2014}{e.g.,}. 
We note that 
its precise value is of small importance, since $\dtcool$ is much larger than $\tmax$ and the tracer expansion time scale. 
The parameters $\ttouch$ and $\tmax$ are determined through a careful inspection of the merger simulation.
In particular, $\ttouch = 12.5 \, {\rm ms}$ is the time where the total entropy starts to increase due to the NS collision
and $\tmax=17.5 \, {\rm ms}$ when it reaches an almost stationary value.
Similarly, for the neutrino mean energies we assume an initial linear increase, followed by an almost stationary phase:
\begin{equation}
\langle E_{\nu} \rangle (t) =
 \left\{
 \begin{array}{ll}
  \Enumin & {\rm for} \quad t \leq \ttouch, \\
  \Enumin +  \langle \Delta E_{\nu} \rangle \, \left( \frac{t - \ttouch}{\tmax-\ttouch} \right) & {\rm for } \quad  \ttouch < t \leq \tmax,  \\
  \Enumax & {\rm for } \quad t > \tmax \, ,
 \end{array}
 \right.
 \label{eq: analytic_meanE_evolution}
\end{equation}
where $\langle \Delta E_{\nu} \rangle \equiv \left( \Enumax-\Enumin \right) $.

Luminosities and mean energies are used to compute local neutrino fluxes $\F_{\nu}$. Far from the neutrino emission region,
we expect purely radial fluxes, axisymmetric around the rotational axis of the remnant.
We further assume a quadratic dependence on $\cos{\theta}$:
\begin{equation}
 \F_{\nu}(R,\theta,t) =  \frac{3 \left( 1 + \alpha \cos^2{\theta} \right) }{3+ \alpha}  \frac{L_{\nu}(t)}{4 \pi R^2 \, \langle E_{\nu} \rangle (t)} \, ,
 \label{eq: neutrino fluxes} 
\end{equation}
where $\theta$ is the polar angle from the rotational axis and $R$ the distance from the remnant center.
For $\alpha = 0$, we recover the isotropic case. For $\alpha = 2$, we mimic the modulation 
of the flux due to the presence of the optically thick disk. In fact, along the equator ($\theta = \pi/2$) 
$\F_{\nu}$ has its minimum, while along the poles ($\theta = 0$) it reaches its maximum.
Numerical results of the neutrino emission from merger remnants point to 
$\F_{\nu}(R,\theta=0,t) / \F_{\nu}(R,\theta=\pi/2,t) \approx 3$ \cite{Dessart.etal:2009,Perego.etal:2014} and
are well described by $\alpha \approx 2$.

The original simulation does not model the neutrino emission. Thus, 
we compute the neutrino properties associated with the merger remnant in a post-processing step.
We map the outcome of a general relativistic binary merger simulation\footnote{Since the 3D distribution of $Y_e$ was not saved for the simulation
presented in the previous section, we computed the neutrino emission
from the output of an almost identical simulation, in which the initial NS are irrotational.}
into the \texttt{FISH+ASL} code \cite{Kaeppeli.etal:2011,Perego.etal:2016}.
The latter has been extensively employed to 
study the evolution of binary merger remnants and their neutrino emission \cite{Perego.etal:2014,Martin.etal:2015}.
For the mapping, we choose a time step corresponding to $\sim 15$ ms after the first touch $\ttouch$. At this time, 
the remnant is characterized by an approximately axisymmetric massive neutron star, surrounded by a thick accretion disk.
Since \texttt{FISH} is a Newtonian Cartesian hydrodynamical code, a further approximation 
becomes necessary because the geometry of the spacetime in a NS is strongly non-Euclidean.
We chose a mapping
as follows: first, we approximate the spacetime as spherically symmetric and extract the 
corresponding metric coefficients from the available 1D data along the $x$ axis. Then 
the volume element after transforming back to Cartesian coordinates becomes unity. 
This way, volume integrals in the Newtonian geometry (in particular the total mass) yield 
approximately the correct GR result, while distances are distorted.

\begin{figure}
 \begin{center}
 \includegraphics[width = 0.5 \linewidth]{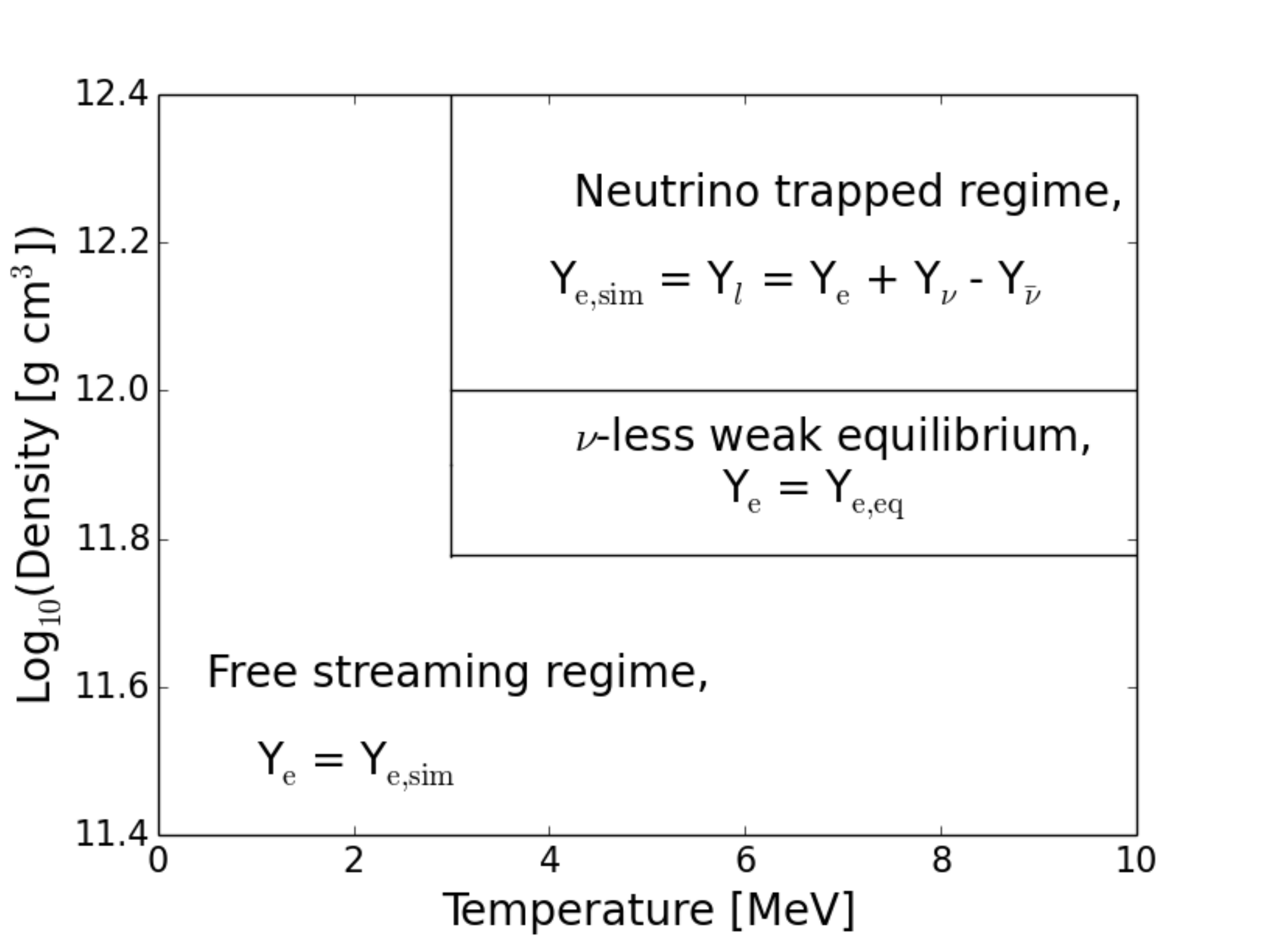}  
 \end{center}
 \caption{Summary of the different regimes adopted to set the initial $\Ye$ in the calculation of the neutrino 
 luminosities and mean energies. 
 The electron fraction obtained in the original simulation is possibly changed according to the thermodynamical conditions 
 inside the remnant. After the initialization, $\Ye$ evolves according to the leakage prescription. 
 See the text for more details.}
 \label{fig:nu_conditions}
\end{figure}

In the original simulation, the initial cold ($T=0$) beta-equilibrium electron fraction is simply advected. 
However, weak reactions at high temperatures and densities are expected to change the 
relative amount of neutrons and protons, as a results of an asymmetric behavior
of electron neutrinos and antineutrinos. In particular, $\Lanue > \Lnue$ \cite{Eichler.etal:1989,Ruffert.etal:1997,Rosswog.Liebendoerfer:2003} 
and the formation of an excess of $\anue$ deep inside the hot remnant \cite{Foucart.etal:2015}
increase the proton and electron abundances.
To take these effects into account, we distinguish between different regimes (also summarized in
Figure~\ref{fig:nu_conditions}) to initialize $\Ye$:
\begin{itemize}
 \item The neutrino trapped regime; due to both high density $\rho > \rho_{\rm eq} = 10^{12}\,{\rm g \, cm^{-3}}$ 
 and temperature $T \gtrsim 3\,{\rm MeV}$, neutrinos diffuse on a time scale longer than the dynamical 
 time scale. Thus, neutrinos are trapped and the electron fraction obtained from the simulation is assumed to be equal 
 to the total lepton fraction, $(\Ye)_{\rm sim} = Y_{\rm l} = \Ye + Y_{\nue} - Y_{\anue}$. The actual values for $\Ye$,
 $Y_{\nue}$, and $Y_{\anue}$ are computed assuming weak equilibrium.
 \item The hot, neutrinoless beta-equilibrium regime; for 
 $\rhofreezeout = 5 \cdot 10^{11}\,{\rm g \, cm^{-3}} < \rho < \rho_{\rm eq} = 10^{12}\,{\rm g \, cm^{-3}}$,
 and $T \gtrsim 3\,{\rm MeV}$, neutrino reactions are still fast enough to change $\Ye$ on a very short time scale 
 ($\lesssim 1\,{\rm ms}$), but neutrino diffusion happens on the same time scale. Under these conditions, we initialize
 $\Ye$ assuming hot, neutrinoless beta-equilibrium.
 \item The neutrino free-streaming regime. For $\rho < \rhofreezeout $ or low matter temperature, the locally produced 
 neutrinos can stream away. For this region, we initially assume that $\Ye = (\Ye)_{\rm sim}$.
\end{itemize}

\begin{figure}
 \includegraphics[width = 0.48 \linewidth]{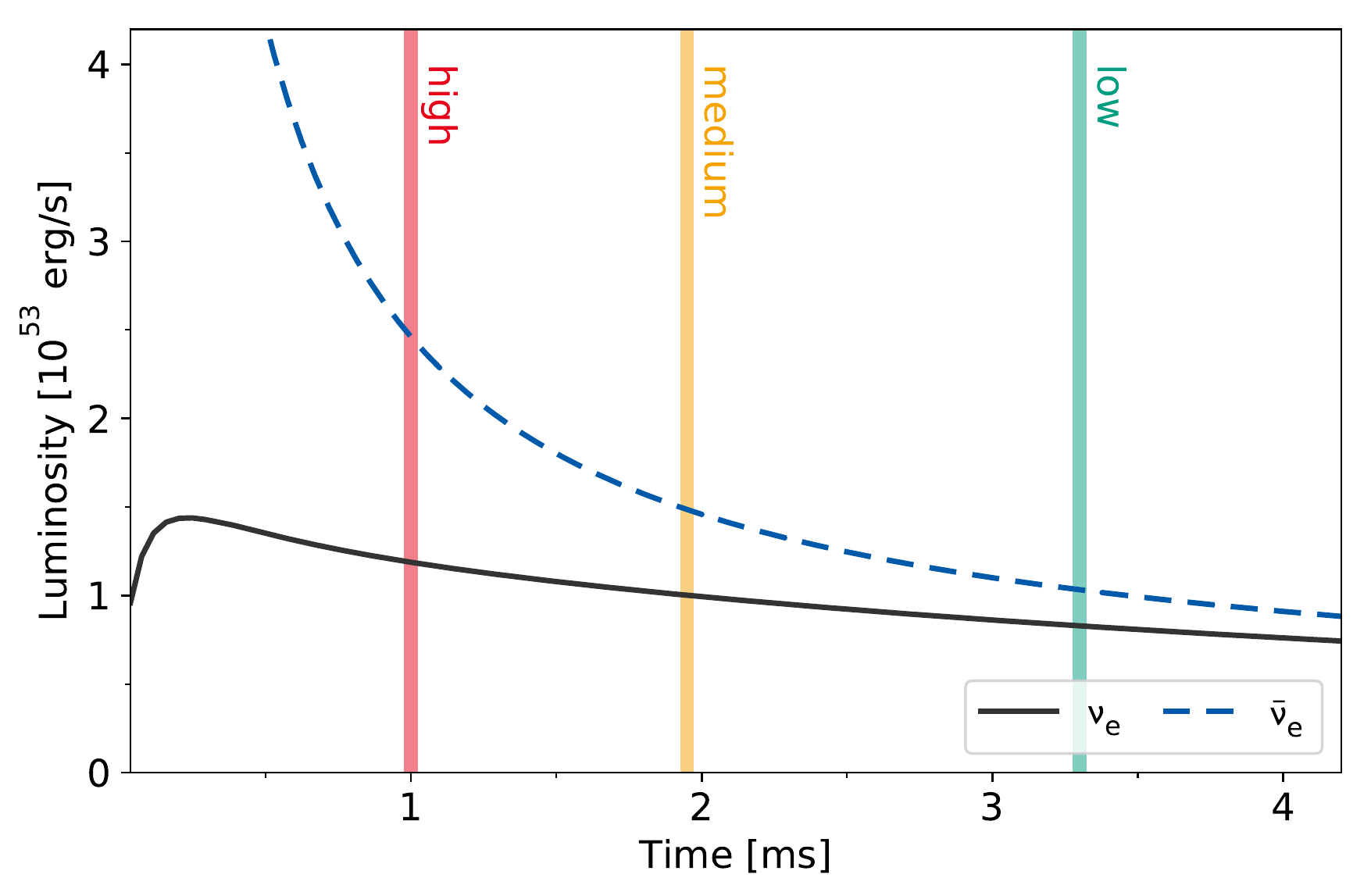}
 \includegraphics[width = 0.48 \linewidth]{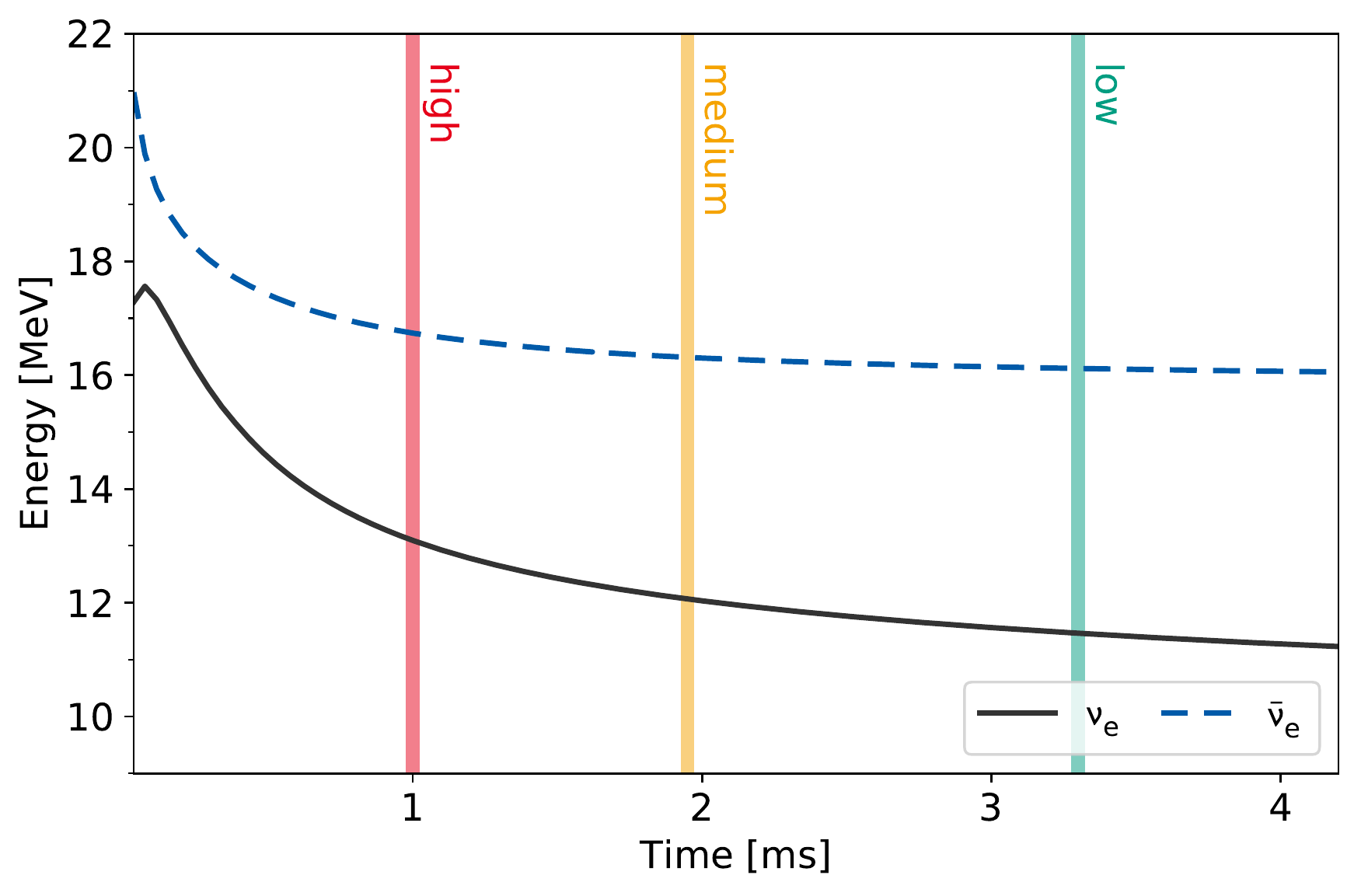} 
 \caption{Temporal profiles of the neutrino luminosities (left) and mean energies (right) obtained by
 post-processing with the \texttt{FISH+ASL} code a snapshot of three dimensional results of the the GR hydrodynamics 
 simulation described in Sec.~\ref{sect:simulation}, once the disk has formed. 
 Neutrino cooling feedback is included only in the temperature 
 and electron fraction evolutions, while matter is considered as a stationary background. Thus, these temporal profiles 
 should be intended as a sequence of possible weak equilibrium configurations (after an initial transient phase),
 rather than a temporal evolution. Vertical bands represent three possible configuration, chosen to bracket uncertainties
 in the determination of the neutrino properties immediately after the merger.}
 \label{fig:neutrino_properties_remnant}
\end{figure}

Due to the differences between the original GR simulation and the Newtonian character of the \texttt{FISH} code, 
we do not dynamically evolve the remnant, but we consider 
its distribution of matter as a stationary background, and we evolve only the electron fraction and the temperature due 
to the neutrino emission. In Figure~\ref{fig:neutrino_properties_remnant}, we
present the temporal evolution of the electron neutrino and antineutrino luminosities (left panel) and mean energies
(right panel), obtained once the simulation has been started. We stress that these temporal profiles do not represent
the true physical evolution of the neutrino quantities, since these can be obtained only by a consistent
radiation-hydrodynamical model. They rather represent a fast approach to ($T,\Ye$) quasi-equilibrium configurations
from an initial non-equilibrium state. In particular, during the first ms, the excess of neutrons in the free streaming 
regime produces a large $\Lanue$ (several $10^{53}\,{\rm erg \, s^{-1}}$), which rapidly changes the electron fraction for $\rho \lesssim \rhofreezeout $. 
This determines a sudden decrease of $\Lanue$. After $\sim 1 \, {\rm ms}$, the electron fraction in the remnant 
has settled to a steady configuration and the luminosities decrease smoothly, due to the remnant and disk cooling.
We use the obtained temporal profiles of the luminosities and mean energies to select three different sets of values,
(referred to as high, medium and low), to span the uncertainties in the determination of the neutrino properties, as well as 
the large range of values obtained in different models. For $\Enumin$, we take 8\,MeV for both $\nue$ and $\anue$.
In Tab.~\ref{tab: lumin-stats}, we summarize all the parameters later used in Eqs.~(\ref{eq: analytic_lum_evolution}) 
and (\ref{eq: analytic_meanE_evolution}). 
We have also tested that the behavior of the luminosities after 1~ms is only weakly dependent 
from the detailed choice of the boundary values (e.g., $\rho_{\rm eq}$ and $\rhofreezeout$) 
in the $\Ye$ initialization.

\begin{table}[!htb]
 \begin{center}
  \caption[Parameters for the maximum neutrino and antineutrino luminosities and energies]
   {Parameters for the maximum neutrino and antineutrino luminosities and energies. 
   In the beginning, we assume vanishing luminosities \mbox{$L_{\nue,\mathrm{min}} = L_{\anue,\mathrm{min}} = 0$} 
   and constant energies \mbox{$ \langle E_{\nue,\mathrm{min}} \rangle = \langle E_{\anue,\mathrm{min}} \rangle = 8 \, \mathrm{MeV}$}.}
  \label{tab: lumin-stats}
  \vspace*{3mm}
  \begin{tabular}{l......}
  \hline
  \hline
  \multicolumn{1}{c}{Name} &
  \multicolumn{1}{c}{$L_{\nue,\mathrm{max}}$} &
  \multicolumn{1}{c}{$L_{\anue,\mathrm{max}}$} & 
  \multicolumn{1}{c}{$\langle E_{\nue,\mathrm{max}} \rangle $} &
  \multicolumn{1}{c}{$\langle E_{\anue,\mathrm{max}} \rangle$} \vspace*{3px} \\
 & \multicolumn{1}{c}{[$10^{53} \, {\rm erg  \, s^{-1}}$]} & \multicolumn{1}{c}{[$10^{53} \, {\rm erg  \, s^{-1}} $]} & \multicolumn{1}{c}{[MeV]} & \multicolumn{1}{c}{[MeV]} \\
  \hline
  capture     & 0.0  & 0.0 & 0.0  & 0.0 \\
  low         & 0.86 & 1.0 & 11.5 & 16.2 \\
  medium      & 1.0  & 1.5 & 12.0 & 16.3 \\
  high        & 1.2  & 2.4 & 13.0 & 16.7 \\
  \hline
  \end{tabular}
 \end{center}
\end{table}

\section{Coupling weak interactions with tracer evolution}
\label{sect:weak-reactions}

We post-process the tracers obtained from the simulation to include the impact of neutrino emission and 
absorption on the evolution of the electron fraction and of the entropy.

If $\tinit$ and $\tendtr$ denote the starting and the ending point of each tracer,
we solve the following system of coupled ordinary differential equations 
for $\tinit < t \leq \tendtr$:
 \begin{eqnarray}
  \frac{{\rm d}\rho}{{\rm d}t} =
   \left( \frac{{\rm d}\rho}{{\rm d}t} \right)_{\rm hydro}(t) \, , \\
 \frac{{\rm d}\mathbf{x}}{{\rm d}t} =
   \left( \frac{{\rm d}\mathbf{x}}{{\rm d}t} \right)_{\rm hydro}(t) \, , \\
 \frac{{\rm d}\Ye}{{\rm d}t} = \left( \frac{{\rm d}\Ye}{{\rm d}t} \right)_{\nu}(t) \, , \\
 \frac{{\rm d}s}{{\rm d}t} =
   \left( \frac{{\rm d}s}{{\rm d}t} \right)_{\rm hydro}(t) + \left( \frac{{\rm d}s}{{\rm d}t} \right)_{\nu}(t) \, .
 \end{eqnarray}
On the lhs, $\rho$ is the matter density, $\mathbf{x}$ the particle position, $\Ye$ the 
electron fraction and $s$ the matter entropy per baryon. On the rhs, time derivatives labeled by ``hydro'' refer
to the evolution obtained inside the simulations, while the ones label by ``$\nu$'' denote the variation due 
to the interaction with neutrinos.
At $t=\tendtr$, temperatures are usually larger than 10\,GK. We extend the evolution also to $t>\tendtr$ 
up to the point $t=t_{\rm end}$ when the temperature reaches 3\,GK,
assuming homologous expansion without shocks ($({\rm d}s/{\rm d}t)_{\rm hydro}=0$) and constant velocity:
\begin{eqnarray}
  \frac{{\rm d}\rho}{{\rm d}t} =   - 3 \frac{\rho_o}{t} \left( \frac{\tendtr}{t} \right)^3 \, , \\
 \frac{{\rm d}\mathbf{x}}{{\rm d}t} =
   \left( \frac{{\rm d}\mathbf{x}}{{\rm d}t} \right)_{\rm hydro}(\tendtr) \, , \\
 \frac{{\rm d}\Ye}{{\rm d}t} = \left( \frac{{\rm d}\Ye}{{\rm d}t} \right)_{\nu}(t) \, , \\
 \frac{{\rm d}s}{{\rm d}t} = \left( \frac{{\rm d}s}{{\rm d}t} \right)_{\nu}(t) \, .
\end{eqnarray}
We note that this expansion is identical to the one used inside the network to evolve the tracers for much
longer time. This ensures a smooth transition between the tracer post-processing and the nuclear network calculations.
We have chosen 3~GK as a limiting temperature because for all our tracers it is below the temperature where the network 
starts to compute detailed abundances out of NSE. We have verified that our results are independent from this choice.

To compute the variations due to neutrinos, $\left( {\rm d}\Ye/{\rm d}t \right)_{\nu}$ and $\left({\rm d}s/{\rm d}t \right)_{\nu}$ , 
we consider a subset of reactions comprising the most relevant charged-current reactions 
between neutrinos and matter, namely the capture of electron, positron, electron neutrinos and antineutrinos on free nucleons:
\begin{eqnarray}
 \p + \el^- \rightarrow \n + \nue \, , \\
 \n + \el^+ \rightarrow \p + \anue \, , \\ 
 \n + \nue \rightarrow \p + \el^- \, , \\ 
 \p + \anue \rightarrow \n + \el^+ \, .
\end{eqnarray}
For each capture reaction, we compute the associated reaction rates 
$\lambda_x$ for species $x=\e^-,\e^+,\nue,\anue$ and
we distinguish between particle ($\lambda_x^0$) and energy ($\lambda_x^1$) rates.
The variation for the electron fraction is
\begin{equation}
 \left( \frac{{\rm d} \Ye}{{\rm d} t} \right)_{\nu} 
 = \left( \lambda^0_{\nue} + \lambda^0_{\el^+} \right) \Yn - \left( \lambda^0_{\anue} + \lambda^0_{\el^-} \right) 
 \Yp \equiv \lambda^0_+ \Yn - \lambda^0_- \Yp \, ,
 \label{eq:ye-evolution} 
\end{equation}
where $\Yn$ and $\Yp$ are the abundances of free neutrons and protons, respectively.

For the entropy variation, from the first principle of thermodynamics we obtain
\begin{equation}
 \left( \frac{{\rm d} s}{{\rm d} t} \right)_{\nu} = 
 \frac{1}{T} \left[ \left( \frac{{\rm d} Q}{{\rm d} t} \right)_{\nu} - \left( \mu_\el-\mu_\n+\mu_\p \right) \left( \frac{{\rm d} \Ye}{{\rm d} t}\right)_{\nu} \right] \, ,
 \label{eq:s-evolution} 
\end{equation}
where $\left( {{\rm d} Q}/{\rm d}t \right)_{\nu} $ is the heat variation due to the emission and absorption of neutrinos:
\begin{equation}
\left( \frac{{\rm d} Q}{{\rm d} t} \right)_{\nu} = 
\left( \lambda^1_{\nue}  - \lambda^1_{\el^+}  \right) \Yn + 
\left( \lambda^1_{\anue} - \lambda^1_{\el^-}  \right) \Yp \, ,
\end{equation}
and $\mu_\el$, $\mu_\p$ and $\mu_\n$ are the chemical potentials of electrons, protons and neutrons, respectively.
The particle and energy capture rates are computed according 
to \cite{Bruenn:1985}, including the corrections due to the electron mass, $\mathcal{M}$, and 
to the weak magnetism $\mathcal{R}_{\nue,\anue}$:
\begin{eqnarray}
\lambda^k_{e^-} & = & \frac{4\pi \sigma_0 c}{(2 \pi \hbar c)^3} \int_0^\infty 
 \left( \frac{E + \Delta}{\me} \right)^2 \, \mathcal{M}(E + \Delta) \,
 \mathcal{R}_{\nue}(E) \, f_{\e^-}(E+\Delta) \nonumber \\ 
 & & \qquad \times E^{2+k} \, \mathrm d E \, , \label{eq:rate-electron-capture}\\
\lambda^k_{e^+} & = & \frac{4\pi \sigma_0 c}{(2 \pi \hbar c)^3} \int_{\Delta + \me}^\infty 
 \left( \frac{E - \Delta}{\me}\right)^2 \, 
\mathcal{M}(E - \Delta) \, \mathcal{R}_{\anue}(E) \, f_{\e^+}(E-\Delta) \nonumber \\ 
 & & \qquad \times E^{2+k} \, \mathrm d E \, , \label{eq:rate-positron-capture}\\
\lambda^k_{\nue} & = &  \frac{\mathcal{G}_{\nue} \sigma_0 c}{(2 \pi \hbar c)^3} \int_0^\infty 
 \left(\frac{E + \Delta}{\me}\right)^2 \,
 \mathcal{M}(E + \Delta) \, \mathcal{R}_{\nue}(E) \, \left[ 1 - f_{\e^-}(E+\Delta) \right] \nonumber \\
 & & \qquad \times  f_{\nue}(E) \, E^{2+k} \, \mathrm d E \, ,  \label{eq:rate-nue-capture}\\
\lambda^k_{\anue} & = & \frac{\mathcal{G}_{\anue} \sigma_0 c}{(2 \pi \hbar c)^3} \int_{\Delta + \me}^\infty 
 \left(\frac{E - \Delta}{\me}\right)^2 
 \mathcal{M}(E - \Delta) \, \mathcal{R}_{\anue}(E) \, \left[ 1 - f_{\e^+}(E-\Delta) \right] \nonumber \\ 
 & & \qquad \times f_{\anue}(E) \, E^{2+k} \, \mathrm d E \, \label{eq:rate-anue-capture},
\end{eqnarray}
where $c$ is the speed of light, $\me$ the electron mass,  $\Delta = 1.2935 \, {\rm MeV}$ 
the mass difference between neutron and proton, and 
$\sigma_0 = 4 (\me c^2)^2 G_F^2 (c_v^2 + 3 c_a^2)/ \pi \hbar^4 \approx 2.43 \times 10^{-44} \, {\rm cm^2}$ 
with $G_F$ the Fermi constant, $\hbar$ the reduced Planck constant, $c_v=1$ and $c_a=g_a \approx 1.23$. 
The distribution functions of electrons and positrons, $f_{\el^\mp}$, are assumed to obey Fermi-Dirac
distributions with non-vanishing chemical potentials.
The electron mass correction term is $\mathcal{M}(x) = \left( 1 - \left( \me / x \right)^2 \right)^{1/2} $,
while the weak magnetism factors, $\mathcal{R}_{\nue,\anue}$, are implemented according to \citeasnoun{Horowitz:2002}.
Their detailed expressions are provided in \ref{ap:weak_magn_corr_and_approx_rates}.
Free-streaming neutrinos and antineutrinos are assumed to be described by a distribution function 
with a Fermi-Dirac energy spectrum of temperature $T_{\nu}$ and zero degeneracy, and with an 
angular dependence $g_{\nu}$,
\begin{equation}
 f_\nu (E,\Omega) = g_{\nu}(\Omega) \, \frac{1}{1 + \exp\left( \frac{E}{\kB T_{\nu}} \right)} \, ,
\end{equation}
such that
\begin{equation}
 \mathcal{G}_{\nu} = \int_{\Omega} \: g_{\nu}(\Omega) \mathrm d \Omega \, .
\end{equation}
The value of $\mathcal{G}_{\nu}$ can be expressed in terms of the local neutrino density and, ultimately, of
the neutrino luminosity and mean energy:
\begin{equation}
  \mathcal{G}_{\nu} = \frac{L_{\nu}}{4 \pi r^2 \left< E_{\nu} \right> c} \frac{(2 \pi \hbar c)^3}{(\kB T_{\nu})^3 F_2(0)} \, ,
\end{equation}
where $F_k(\eta) \equiv \int_0^\infty x^k/[1+\exp(x-\eta)] \, {\rm d}x$ is the Fermi integral of order $k$ evaluated at $\eta$. 
In the rates calculations, we have included Pauli blocking factors for electrons and positrons in the final
states, while we have neglected neutrino blocking factors in free streaming conditions.

Hot and dense matter in NSE is described by a nuclear equation of state in tabular
form \cite{Hempel.etal:2012}. For consistency with the underlying simulation, we choose the NL3 parameterization for the 
nucleon interaction. However, since our initial densities are usually below $10^{13} \, {\rm g \, cm^{-3}}$, we 
do not expect our choice to have a significant impact on the results.

For each trajectory, we assume as initial conditions for the matter density, entropy, and velocity
the tracer properties at $t=\tinit$, such that $\rho(\tinit) = 10^{12} \, {\rm g \, cm^{-3}} \equiv \rho_{\rm eq}$. We consider 
this value for $\rho$ as the transition between the diffusive and the free-streaming regime 
for neutrinos (see Sec.~\ref{sect:weak-reactions}). 
A precise initialization of the electron fraction would require a detailed evolution of $\Ye$
inside the merger simulation in neutrino diffusive conditions. Since this is not available, we consider
two opposite cases:
\begin{itemize}
 \item Case A, in which weak equilibrium is assumed down to $\rho_{\rm eq}$. This corresponds to the case where
 neutrino reactions are fast enough (compared to the expansion time scale) to drive $\Ye$ toward the equilibrium
 value associated with the corresponding density and temperature, $Y_{\rm e,eq}$. If the tracer starts at a density 
 below $\rho_{\rm eq}$\footnote{This condition is fulfilled for $\sim 25 \%$ of the tracers.}, 
 then $\Ye=0.044$ is assumed (this value corresponds to mass-weighted average $\Ye$ of ejecta passing through a
 spherical surface of radius 111~km in the original simulation).
 \item Case B, where $\Ye=0.044$ is assumed for every tracer at $t=\tinit$. This corresponds to the case 
 where all neutrino reactions are very slow for $\rho > \rho_{\rm eq}$, compared to the expansion time scale, 
 and the electron fraction remains practically unchanged.
\end{itemize}
For most of the tracers, temperatures are usually smaller than $0.5 \, {\rm MeV}$ for $\rho \approx \rho_{\rm eq}$. 
In addition, NSE predicts a composition characterized by free neutrons (with mass fraction 
$X_{\rm n} \sim 1 - \Ye $), neutron-rich nuclei ($X_{\rm heavy} \sim \Ye $) and a negligible amount of
free protons, for initially low electron abundances ($\Ye \lesssim 0.2$). Under these conditions, 
neutrino reactions are expected to be slower than the expansion time scale. In particular,
the absorption of electron neutrinos on free neutron can increase $\Ye$, but the rate is usually too
slow to reach $Y_{\rm e,eq}$.
Thus, we expect our two cases to bracket the actual evolution of the electron fraction.

\begin{figure}
 \centering
 \includegraphics[width = 0.49\linewidth]{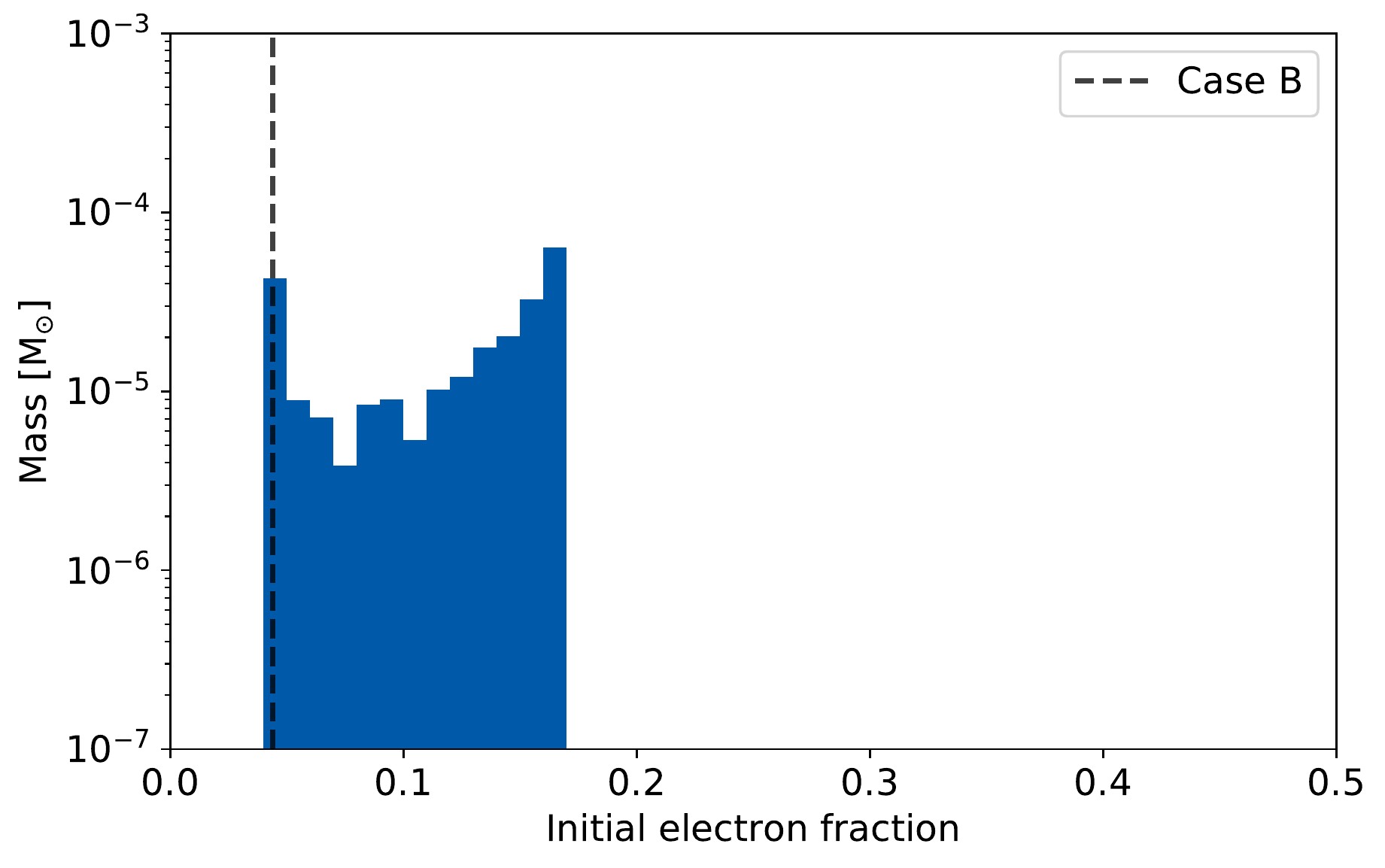}
 \caption{Mass distribution of the electron fraction of the ejecta at the beginning of the post-processing calculations
 for case A, (i.e., assuming weak equilibrium up to $t_{\rm init}$ when matter density drops below $\rho_{\rm eq}$). 
 Case B corresponds to the vertical line at $\Ye=0.044$.}
 \label{fig:initial_ye_distribution}
\end{figure}

Figure~\ref{fig:initial_ye_distribution} shows the mass distribution 
of the electron fractions for all tracers at $\tinit$, for case A. 
Due to the initial high densities and low temperatures of the tracers, the average $\Ye$ is rather low 
($\langle \Ye \rangle \approx 0.12$) and close to 
the values obtained in the simulation without the inclusion of weak reactions.

\subsection{Nuclear network}
In order to determine the composition of the fluid along the
trajectories, we employ a complete nuclear reaction network
\cite{Winteler.etal:2012,Korobkin.etal:2012,Martin.etal:2015}.
It includes over 5800 nuclei between the valley of stability and the neutron drip line, comprising isotopes from H to Rg. 
The nuclear properties (e.g., mass excess, ground state spin, and partition functions) and reaction rates are taken 
from the compilation of \citeasnoun{Rauscher.Thielemann:2000} and \citeasnoun{Rauscher:2003} for the Finite Range Droplet Model 
\citeaffixed{Moeller.etal:1995}{FRDM, }. In particular, the reaction rates are tabulated as the coefficients of a fit function 
in the JINA REACLIB format \cite{Cyburt.etal:2010}. Theoretical weak interaction rates including neutrino absorption 
on nucleons are taken into account \cite{Fuller.etal:1982a,Fuller.etal:1982b,Fuller.etal:1985,Langanke.MartinezPinedo:2001,Moeller.etal:2003}, 
and to compute them we utilize chemical potentials from the Helmholtz equation of state \cite{Timmes.Swesty:2000}. 
Furthermore, neutron capture for nuclei with $Z > 83$ and neutron-induced fission rates are taken from \citeasnoun{Panov.etal:2010} 
while beta-delayed fission probabilities are from \citeasnoun{Panov.etal:2005}.

We feed the reaction network with the temporal profiles of the matter density 
and radial position obtained by post-processing the  ejected tracers, see Sec.~\ref{sect:weak-reactions}. 
The electron fraction, temperature, and nuclear composition are only initialized and then evolved consistently 
by the network. This ensures a smooth transition between the two different post-processing steps. 
As starting point of the $\Ye$ and nucleosynthesis calculations, we still consider NSE conditions 
occurring at $T \approx 8$~GK. These conditions typically occur a few tens of milliseconds
after the shock has reheated the outflow. From then on, we switch to the full 
reaction network to determine the nucleosynthesis, while descending to lower 
temperatures and densities. For temperatures below 3~GK, we further extrapolate 
the expansion inside the network assuming a homologous outflow.
The energy generation by the r-process is calculated and its impact on
the entropy is included \cite{Freiburghaus.etal:1999a,Korobkin.etal:2012} 
The heating mainly originates from beta decays and we assume that the energy 
is roughly equally distributed between thermalizing electrons and photons, 
and escaping neutrinos and photons \citeaffixed{Metzger.etal:2010b,Barnes.etal:2016}{see}{ for a recent discussion}. 
We compute the final abundances at $10^9$\,years.

\section{Results}
\label{sect:results}

\subsection{Representative tracers}
\label{sec: representative tracers}

Both the shock and the neutrino irradiation have a strong effect on the electron fraction evolution. 
In the following, we examine how the considered reactions influence the electron fraction in detail. 

\begin{figure}[!htb]
 \centering
 \includegraphics[width=0.49\linewidth]{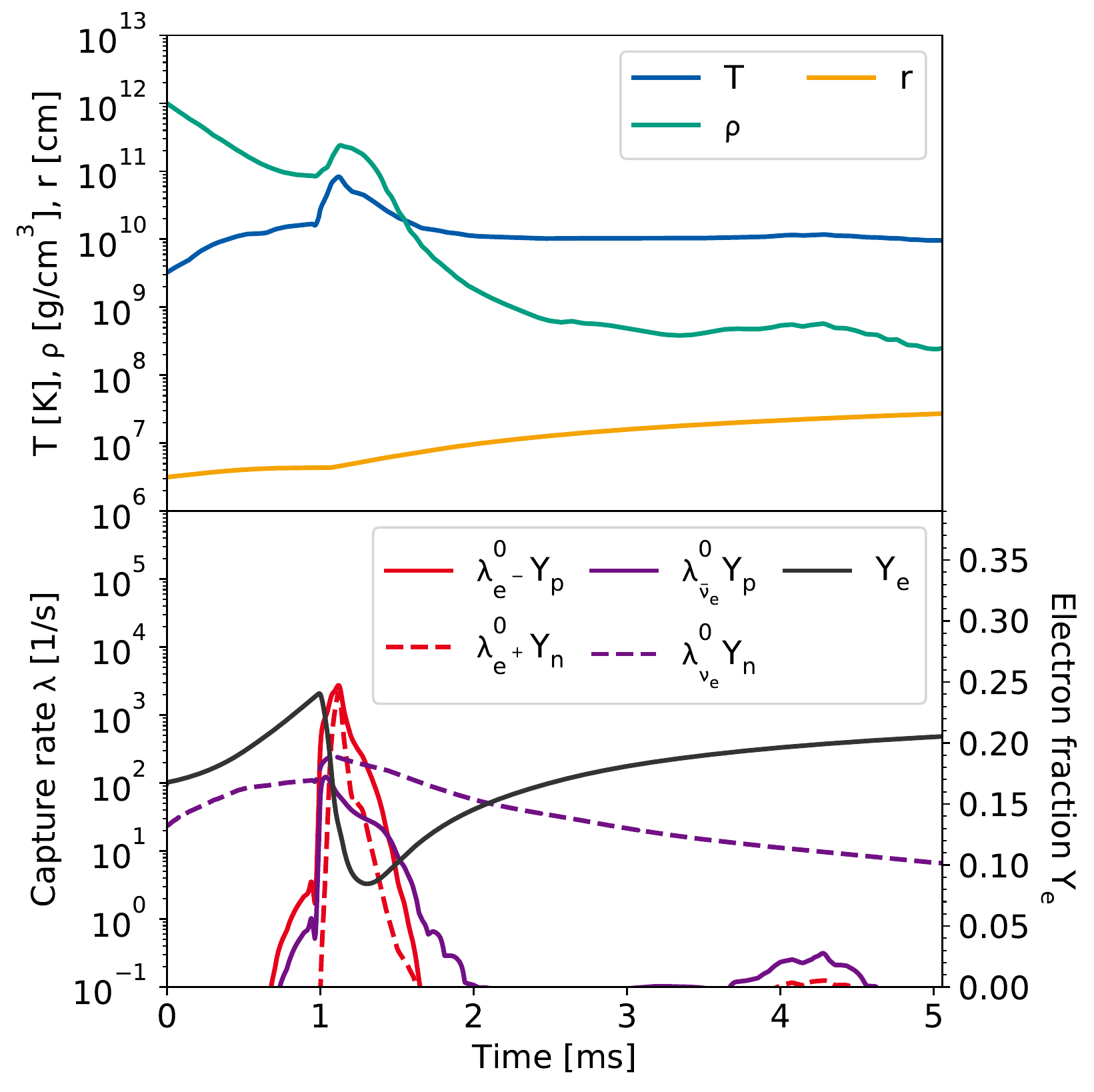}
 \includegraphics[width=0.49\linewidth]{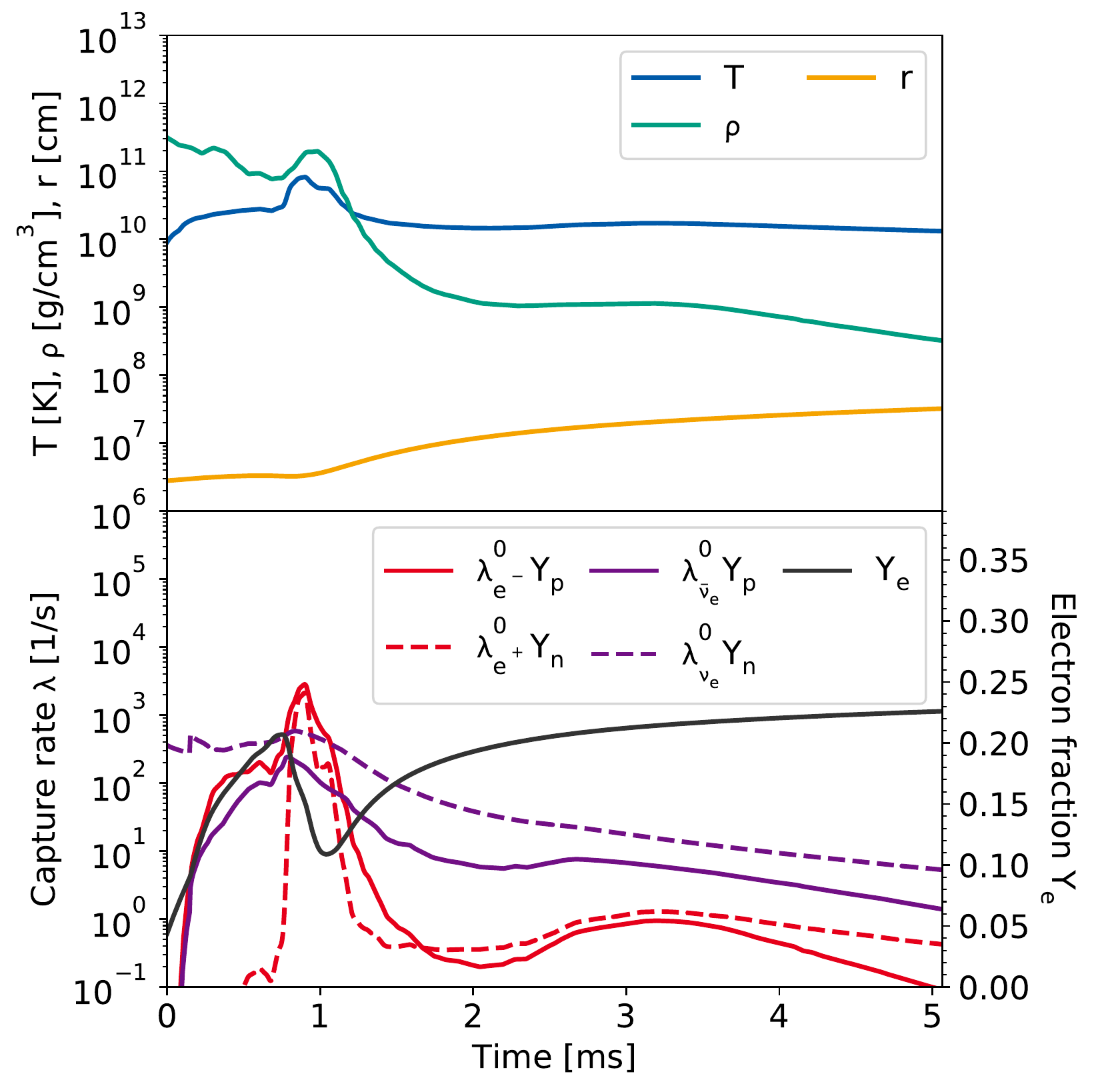}
 \caption{Temporal evolution for two selected tracers of case A with
   high isotropic luminosities. The left panel shows a tracer with
   $Y_{\e,{\rm ini}} = 0.044$, while the right panel shows a tracer
   starting from beta-equilibrium conditions with $Y_{\e,{\rm ini}}
   \sim 0.16$. Top panels: hydrodynamic properties of a representative
   tracer. Bottom panels: trends of the considered reaction rates and
   evolution of the electron fraction over time. Rates that lead to a
   decrease in the electron fraction are shown with solid lines, while
   rates that cause an increase of the electron fraction are plotted
   with dashed lines.}
 \label{fig:evolution-hydro-rates-ye}
\end{figure}

In the upper panels of Figure~\ref{fig:evolution-hydro-rates-ye}, we show the hydrodynamical properties of two representative tracers, initialized 
according to case A.
One tracer (left panel) starts with a density of $10^{12}\,{\rm g \, cm^{-3}}$ and an initial weak equilibrium electron fraction of $0.16$. 
In the other tracer (right panel), the initial density is below $10^{12}\,{\rm g \, cm^{-3}}$ and the initial $\Ye$ is assumed to be 0.044 
(see Sec.~\ref{sect:weak-reactions}).
The former tracer is also representative of the trajectories initialized according to case B. A more extensive discussion 
about the differences between case
A and B will be provided in Sec.~\ref{sec: property distributions}.
The lower panels illustrate the reaction rates calculated with Eqs.~(\ref{eq:rate-electron-capture})$-$(\ref{eq:rate-anue-capture}) as well as 
the resulting evolution of the electron fraction as a function of time. When the tracer leaves the neutrino diffusion regime, 
the relatively low temperatures favor the formation of neutron-rich, tightly bound nuclei 
in combination with free neutrons in NSE. Particularly, the abundance of free protons vanishes under such cold, neutron-rich conditions 
(cf. Sec~\ref{sect:weak-reactions}). 
Electron neutrino absorption on free neutrons increases the electron fraction, despite the non-negligible
effect of Pauli blocking for the degenerate electrons in the final state.
As the density decreases and the temperature increases, nuclei are
more and more dissociated into free nucleons in NSE. After about 1\,ms, 
a shock sets in and the sudden rise of temperature triggers electron captures and, to smaller extent, positron captures. 
Hence, the electron fraction drops sharply. 
As soon as the temperature decreases, as a consequence of the expansion induced by the shock itself, 
the subsequent evolution is mainly determined by 
electron neutrino captures on neutrons.
When material expands to larger distances, the (anti)neutrino fluxes 
fade as $R^{-2}$ and the electron fraction flattens after a few milliseconds. In the subsequent expansion phase, 
density and temperature decrease monotonically, 
until leaving NSE and reaching conditions relevant for the r-process nucleosynthesis.

\begin{figure}[!htb]
 \centering
 \includegraphics[width=0.49\linewidth]{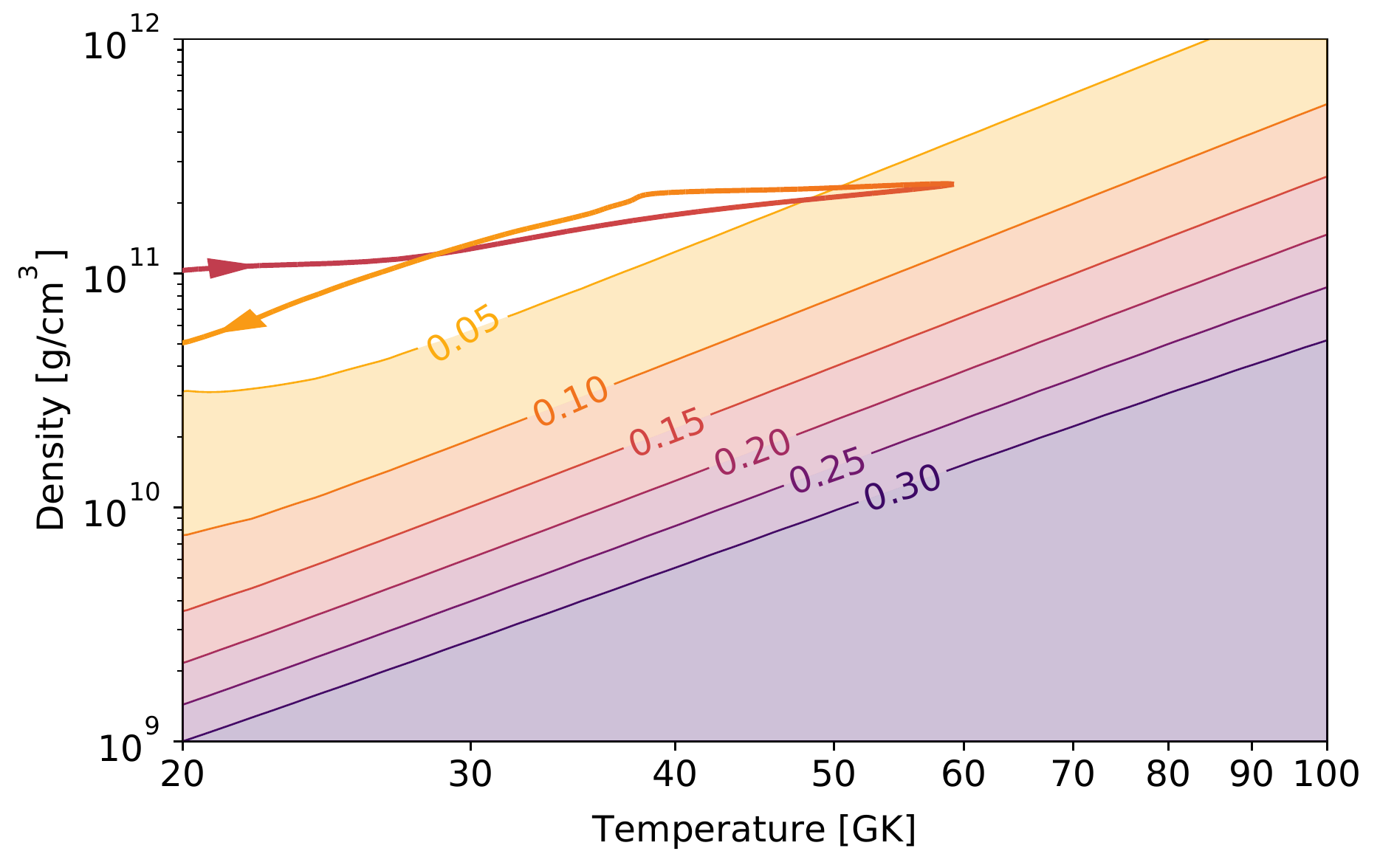}
 \caption{Comparison of the electron and positron captures for different conditions in the density-temperature
 plane. The thick line shows the electron fraction evolution along the representative trajectory 
 from Figure~\ref{fig:evolution-hydro-rates-ye} including only electron and positron captures. 
 Arrows indicated the temporal evolution of the hydrodynamics along the trajectory. The thin lines in the 
 colored area mark critical values of density and temperature, 
 labeled with the corresponding electron fraction in NSE. 
 These are the conditions at which the electron capture rate equals the positron capture rate, 
 i.e., $\lambda^0_{\mathrm{e}^-} Y_\p / \left( \lambda^0_{\mathrm{e}^+} Y_\n \right) = 1$. 
 At higher densities, electron capture dominates, while at lower densities positron captures win. 
 Note that we only show contour lines for electron fractions up to $\Ye = 0.30$.}
 \label{fig:critical-density-temperature}
\end{figure}

It is instructive to discuss the effects of the shock passage on the electron
fraction in terms of the different conditions experienced in the density-temperature plane. 
Since (anti)neutrino capture rates, $\lambda_{\nue}$ and 
$\lambda_{\anue}$, are weakly affected by density and temperature gradients in the free-streaming regime, 
we restrict our discussion to the electron 
and positron captures. Using Eqs.~(\ref{eq:rate-electron-capture})~and~(\ref{eq:rate-positron-capture}), 
we evaluate the rates for a large sample of density and temperature conditions. To present
their impact on the evolution of the electron fraction, we consider the product of the rates and 
the corresponding target abundance, i.e.,
$\lambda_x Y_x$. The nucleon abundances are calculated for NSE conditions with the aid of the nuclear EOS. 
Figure~\ref{fig:critical-density-temperature} shows with contour lines the conditions where 
electron and positron captures balance each other,
\mbox{$\lambda_{\mathrm{e}^-} Y_\p / \left( \lambda_{\mathrm{e}^+} Y_\n \right) = 1$}. The contour lines 
are labeled with the corresponding electron fraction and illustrate a range of $\Ye$ between 0.05 and 0.30 
in steps of $\Delta \Ye = 0.05$. 
For a given $\Ye$, in the region above the corresponding line electron captures dominate, while positron 
captures win for lower densities or larger temperatures.
This plane can be understood in terms of the degree of degeneracy of the electrons. 
Making use of the approximated expressions of the rates derived in~\ref{ap:weak_magn_corr_and_approx_rates}, we obtain:
\begin{equation} \label{eq:ratio-elpos-captures}
 \frac{\lambda^0_{\mathrm{e}^-} \, Y_\p}{\lambda^0_{\mathrm{e}^+} \, Y_\n} \approx 
 \frac{F_4 \left( \eta_\e - \frac{\Delta}{\kB T} \right) \, \Ye}{F_4 \left( - \eta_\e \right) \, \left( 1-\Ye \right)} \, ,
\end{equation}
where $\eta_\e$ is the electron degeneracy parameter, defined as $ \eta_\e \equiv \mu_\mathrm{e} / \kB T$,
$\mu_\e$ the relativistic electron chemical potential including the rest mass.
We have also assumed that temperature is high enough to dissociate most of the nuclei in free nucleons. 
When electrons are degenerate (i.e., for high densities and low temperatures), $ \eta_\e \gg 1$ and 
$F_4 \left( \eta_\e - \Delta/ (\kB T)  \right) / F_4 \left( - \eta_\e \right) 
\sim \left( \eta_\e - \Delta/(\kB T) \right)^5 e^{\eta_\e} /120 \gg 1 $ \cite{Bludman.vanRiper:1977,Takahashi.etal:1978}.
On the contrary, in non-degenerate conditions $ \eta_\e \sim 0$ and
$F_4 \left( \eta_\e - \Delta/ (\kB T)  \right) / F_4 \left( - \eta_\e \right) 
\sim \left[ 1 + 0.974 \left( \eta_\e - \Delta/(\kB T) \right) \right] $.
Thus, for high temperatures and/or low densities such that $\mu_\e \lesssim \Delta $, the positron capture rate becomes dominant. 
More in general, for $\mu_\e \lesssim \Delta/2$ the rates become comparable for all regimes. 

We evolve again the tracer from the left panels of Figure~\ref{fig:evolution-hydro-rates-ye}, but including only electron and positron captures 
in Eqs.~(\ref{eq:ye-evolution}) and (\ref{eq:s-evolution}). We show the subsequent evolution of its $\Ye$ with a colored thick line 
in Figure~\ref{fig:critical-density-temperature}. 
Its color changes according to the evolution of the electron fraction and 
coherently with the thin threshold lines.
When the ejecta are hit by the shock, the electron fraction of the trajectory is close
to the weak-equilibrium value, $\Ye \approx 0.16$, given no (anti)neutrino absorption. 
For the high degeneracy conditions experienced at the shock passage,
electron captures are initially favored by a factor of $\sim 100$ over positron captures. 
As the temperature increases toward the peak value, the electron
degeneracy decreases and this factor goes down, but not enough to make positron captures dominant.
Instead, the ongoing electron captures rapidly decrease the electron fraction, shifting the 
line of $\lambda^0_{\mathrm{e}^-} Y_\p / \left( \lambda^0_{\mathrm{e}^+} Y_\n \right) = 1$ into the direction 
of the conditions found for the trajectory.
This effectively flattens the evolution of the electron fraction, as it tends to balance the capture rates. 
At the shock peak, the reaction rates are fast enough to reach the corresponding weak equilibrium conditions.
When the electron fraction has dropped to $\Ye \approx 0.07$, the ratio of electron and positron capture rates 
is close to unity. However, the temperature is already very low at this point, shutting off both kinds of captures. 
If these reaction types were the only ones involved in the electron fraction evolution, then its profile 
would remain constant at later times. 

\subsection{Property distributions}
\label{sec: property distributions}

\begin{figure}[!htb]
 \centering
 \includegraphics[width=0.95\linewidth]{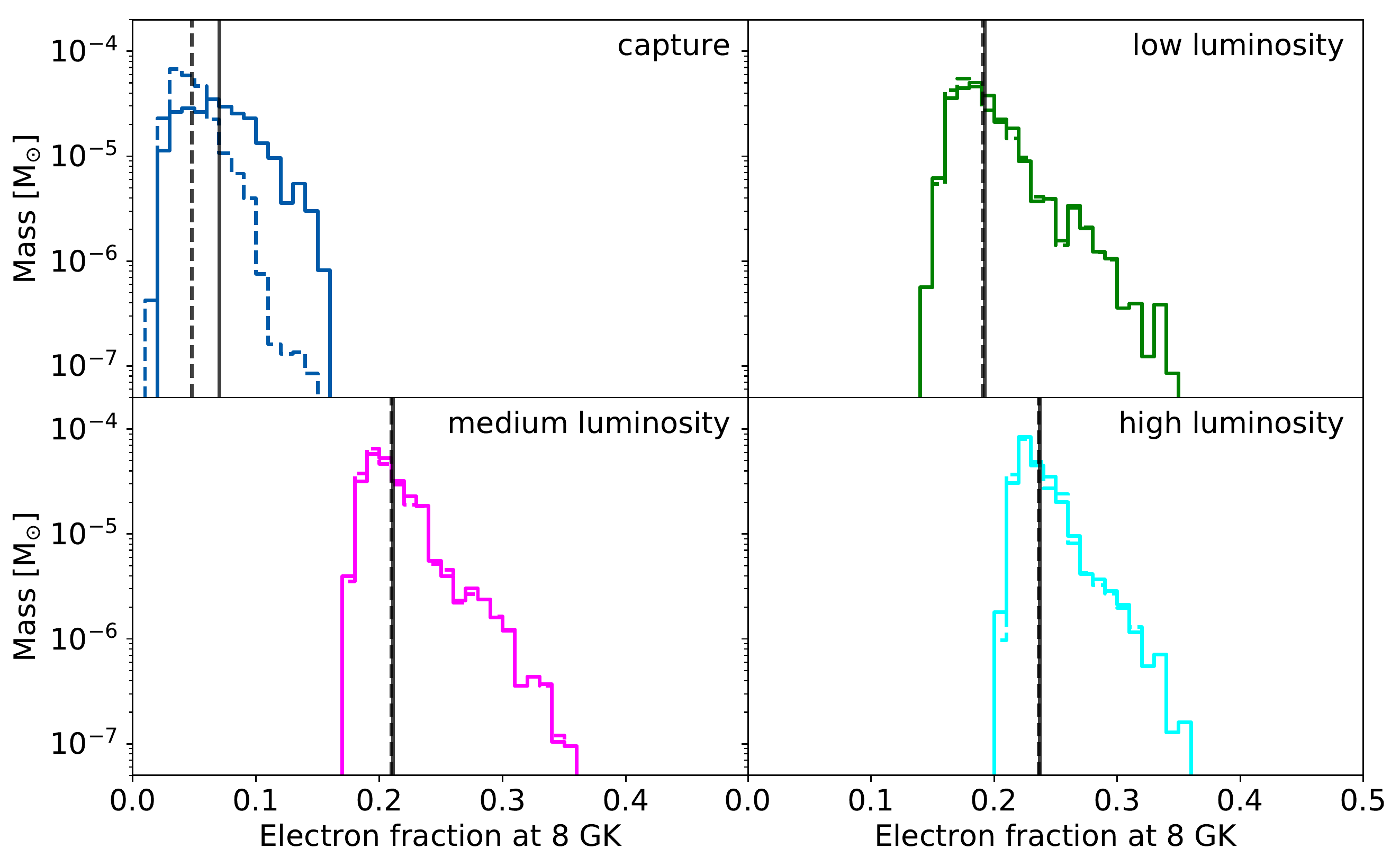}
 \caption{Mass distributions of the electron fraction at 8~GK obtained from the post-processed tracers. 
 Different panels refer to different treatments of the weak reactions. In the top-left panel, we include only electron and positron captures, while 
 in the other panels also (anti)neutrino captures with increasing, isotropic luminosities. Solid and dashed lines refer 
 to the results obtained using case A and case B as initial conditions, respectively. Vertical lines mark the average electron fractions.
 In all cases, we obtain a rather broad 
 distribution of the electron fraction, with generally less neutron-deficient values, as we assume higher (anti)neutrino luminosities.}
 \label{fig:ye-weak-reactions_iso_lum_alpha0}
\end{figure}

\begin{figure}[!htb]
 \centering
 \includegraphics[width=0.95\linewidth]{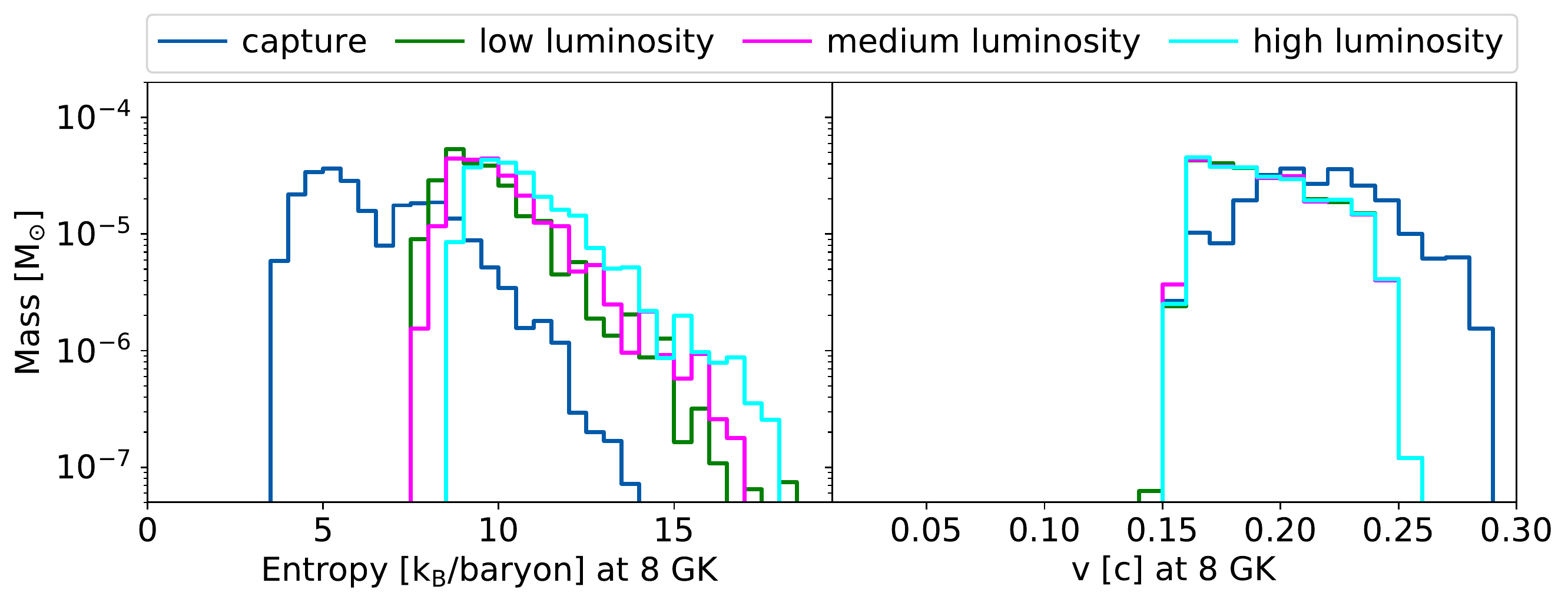}
 \caption{Mass distributions of the entropy (left) and radial velocity (right) at 8~GK, obtained by evolving ejected tracers
 initialized according to case A. Different colors refer to different treatments of the weak reactions, while neutrino 
 luminosities are assumed to be isotropic
 (cf. Figure~\ref{fig:ye-weak-reactions_iso_lum_alpha0}).}
 \label{fig:s_v_iso_lum_alpha0}
\end{figure}

The nucleosynthesis in the ejecta is sensitive to the electron fraction, $\Ye$, the entropy, $s$, and the expansion time scale 
(or equivalently the expansion velocity, $v_{\rm 8GK}$) at NSE freeze-out \citeaffixed{Hoffman.etal:1997}{e.g., }.
Using the whole ensemble of ejected tracers from the simulation, we obtain distributions for all these quantities, recorded
when the tracer temperature drops below 8 GK, well after the shock has passed. At later time, weak processes can 
still change $\Ye$ and $s$. However, due to the low temperatures and large distances from the remnant 
($R \gtrsim 600 \,{\rm km} $), 
these residual variations are small ($\delta \Ye / \Ye \sim 0.02 $ and $\delta s / s \sim 0.015$).

\begin{figure}[!htb]
 \centering
 \includegraphics[width=0.7\linewidth]{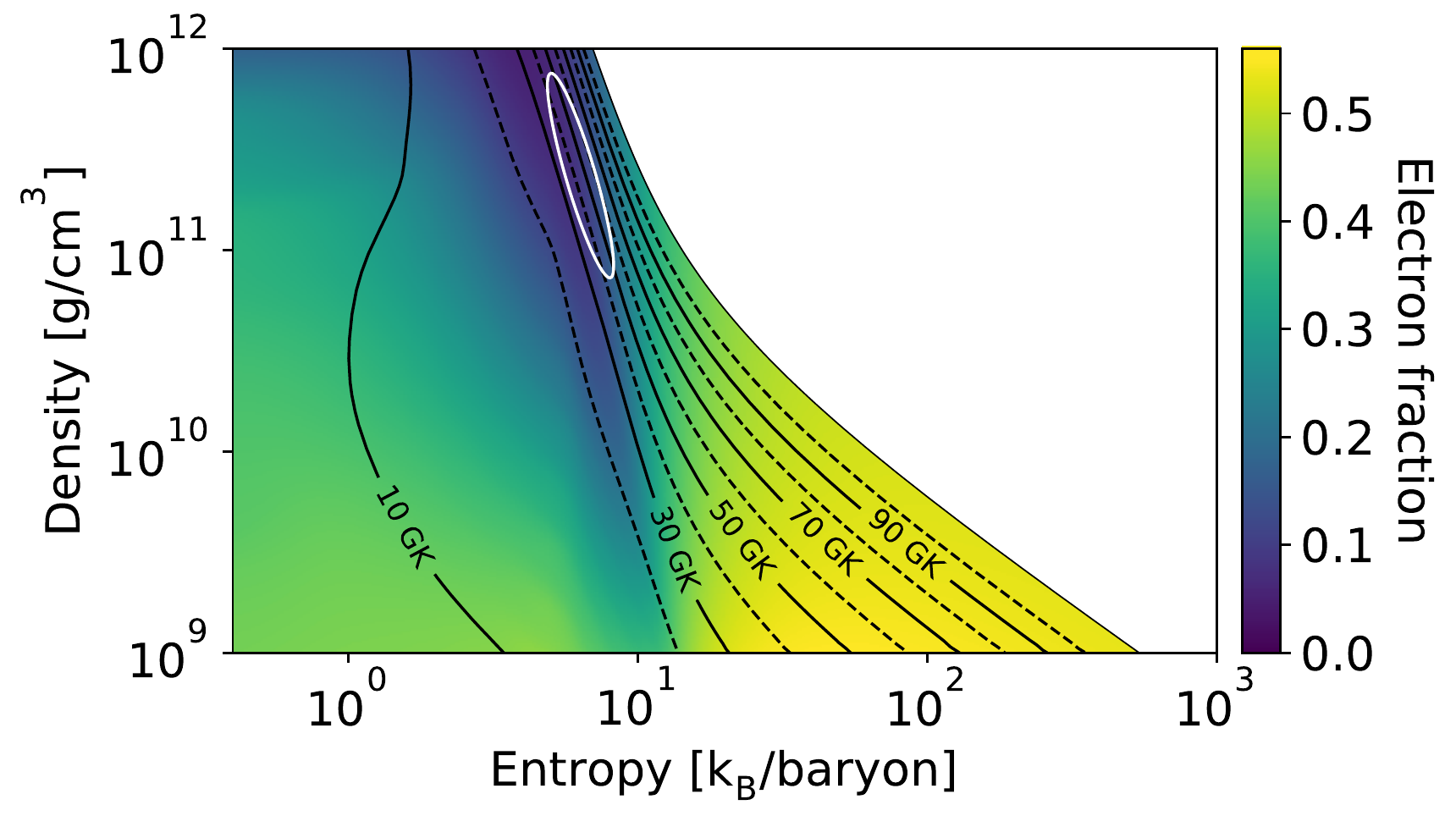}
 \caption{Beta-equilibrium values for the electron fraction in the entropy-density plane. Dashed and solid black lines indicate temperature contours. The white ellipse shows the one-sigma environment of the conditions that the tracers encounter at the peak temperature of the shock. Accordingly, the electron fraction tends to evolve toward an equilibrium value of $\Ye \sim 0.10$ if only electron and positron captures are considered.}
 \label{fig:betaEquilibrium-density-entropy}
\end{figure}

In Figure~\ref{fig:ye-weak-reactions_iso_lum_alpha0}, we present the electron fraction distributions 
for different treatments of the weak reactions (different panels) and for both cases A and B (with solid and dashed lines, respectively).
If we only include electron and positron captures to evolve the tracers 
(top-left panel)\footnote{We notice that this treatment of the weak reactions is equivalent to what is done in simulations employing
leakage schemes without absorption terms in the optically thin regime.},
different initial conditions results in different $\Ye$ distributions.
In particular, the broader and less neutron-rich 
initial distribution assumed in Case A is reflected in the evolved
distribution at 8~GK. A closer comparison reveals that in this case the peak and the mass-weighted average 
are located around $\Ye = 0.07$, thus reduced compared to the beta-equilibrium values obtained 
at weak freeze-out (cf., Figure~\ref{fig:initial_ye_distribution}).
In the absence of neutrino captures, the electron fraction is not significantly modified after being processed 
by the shock wave. This is a direct consequence of the effect of the shock passage discussed in 
Sec.~\ref{sec: representative tracers}.
As soon as the temperature increases due to the shock passage, electron and positron captures are significantly enhanced 
(cf., Figure~\ref{fig:evolution-hydro-rates-ye}). 
In Figure~\ref{fig:betaEquilibrium-density-entropy}, we show the electron fraction in 
neutrinoless beta-equilibrium on the matter density-entropy plane. The white ellipse 
contains the hydrodynamical conditions experienced by most of the tracers at the shock peak
and shows that the resulting equilibrium $\Ye$ scatters around $\Ye \sim 0.10$
for most of the ejecta.
In case B, the lower initial $\Ye$ values ($\sim 0.044$) favor positron captures, which increase the electron fraction 
towards the equilibrium conditions, where $\lambda^0_{\e^-} Y_\p / \left( \lambda^0_{\e^+} Y_\n \right) = 1$. 
However, due to the vicinity of the initial electron fraction
to the equilibrium one, and to the relatively slow positron capture rate 
($\lambda^0_{\e^+} Y_\n \, \Delta t_{\rm shock} \lesssim 1$, where $\Delta t_{\rm shock} \approx 0.5 \, {\rm ms}$ 
is the width of the shock duration), 
the resulting rise in $\Ye$ is only marginal and the electron fraction distribution peaks again around $\Ye=0.05$. 
On the contrary, ejecta in case A start often with a considerably higher weak equilibrium 
electron fraction (see Figure~\ref{fig:initial_ye_distribution}). 
The conditions at the shock peak are relatively far from the $\lambda^0_{\e^-} Y_\p / \left( \lambda^0_{\e^+} Y_\n \right) = 1$ line, 
and electron capture is enhanced by electron degeneracy up to two orders of magnitude with respect to positron capture.
Moreover, weak reactions are fast enough to approach equilibrium, since for most of the 
trajectories $\lambda^0_{\e^-}  Y_\p \, \Delta t_{\rm shock} > 1$.
This leads to an appreciable decrease in the electron fraction, compared to the initial conditions of 
case A, Figure~\ref{fig:initial_ye_distribution},
and at the same time to a qualitative different behavior compared to case B, where equilibrium at the shock is almost never reached.

The other panels of Figure~\ref{fig:ye-weak-reactions_iso_lum_alpha0} show the three cases including (anti)neutrino 
absorption reactions for different strength of the neutrino luminosities (see Table~\ref{tab: lumin-stats}). 
We first consider the case of isotropic neutrino emission (i.e., $\alpha=0$ in Eq.~(\ref{eq: neutrino fluxes})). 
For all considered combinations, the late-time electron fraction is generally shifted to higher values, compared 
to the starting beta-equilibrium values.
This is mainly a consequence of the neutrino absorption occurring after the passage of the shock wave. 
The mass-weighted average is contained within the range $0.19 \lesssim \langle \Ye \rangle \lesssim 0.23$ 
and larger neutrino fluxes result in higher $\langle \Ye \rangle$. The distributions have a tail toward high electron 
fractions, up to $\Ye \approx 0.36$ for all degrees of irradiation.
Once neutrino absorptions are taken into account, the differences between case A and B vanish.
This is due to copious absorptions of electron neutrinos on neutrons, which occur before the shock passage and for both cases A and B. 
The subsequent increase in the electron fraction, 
well above $\Ye \approx 0.10$, sets the tracer conditions similar to the ones we have discussed above, in the case of
electron and positron captures alone and initial conditions in case A. Due to the achievement of the equilibrium $\Ye$ inside the shock,
the differences between the cases A and B disappear and the variety in the final distributions depends only on the degree of neutrino 
irradiation after the shock passage.

Since our results and conclusions are largely independent from the initial conditions, in the following we will consider only case A.
The distributions for the entropy per baryon and for the asymptotic expansion velocity are presented 
in Figure~\ref{fig:s_v_iso_lum_alpha0}.
The passage of the shock wave increases the entropy from $\sim 4~k_{\rm B}~{\rm baryon}^{-1}$ to 
$\sim 8~k_{\rm B}~{\rm baryon}^{-1}$. In the electron-positron capture case, the emission of neutrinos removes efficiently 
entropy from the fluid elements. The resulting distribution at 8 GK shows 
two peaks, one around $5~k_{\rm B}~{\rm baryon}^{-1}$ and one around $7~k_{\rm B}~{\rm baryon}^{-1}$,
which correlate with the peak temperature after the shock passage: the more intense neutrino emission
from the hotter tracers ($T_{\rm peak} > 40~{\rm GK}$) reduces the entropy more significantly than from the 
colder ones.
If neutrino absorption processes are included, the captures of high energy (anti)neutrinos on expanding and 
cooling matter compensate the reduction of entropy provided by
neutrino emission. The bimodal distribution 
observed before is substituted by a distribution with a single peak around 
$8-9~k_{\rm B}~{\rm baryon}^{-1}$ and a rapidly decreasing tail, extending up to $16~k_{\rm B}~{\rm baryon}^{-1}$.
Due to the balance between emission and absorption processes, 
we notice only a marginal increase in the entropy profiles for increasing neutrino luminosities.
The ejecta expand with fast velocities ($v_{\rm 8\,GK} > 0.15~c$) in all cases. The wider and faster distribution 
obtained in the electron-positron capture case is simply a consequence of the more rapid cooling observed in this case. 
Once neutrino absorption processes are taken into account, the NSE freeze-out temperature $T_{\rm NSE}=8\,\mathrm{GK}$) is reached at later times 
and larger radii. In these cases, the radial velocity of each fluid element has further 
decreased due to the motion inside the gravitational well, and has approached its asymptotic value, $v_{\infty}$.

\begin{figure}[!htb]
 \centering
 \includegraphics[width=0.95\linewidth]{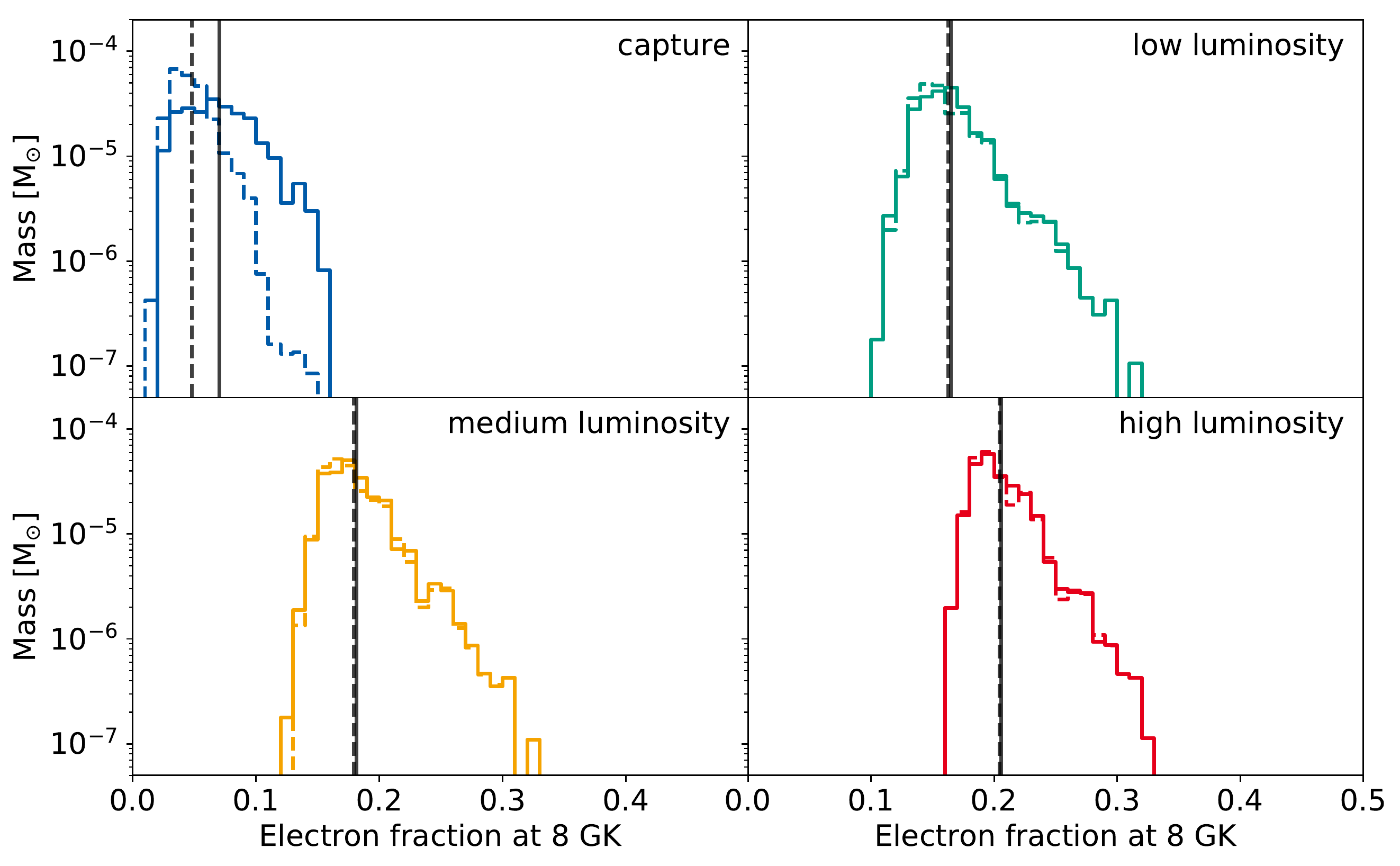}
 \caption{Same as in Figure~\ref{fig:ye-weak-reactions_iso_lum_alpha0}, but for anisotropic neutrino luminosities.}
 \label{fig:ye-weak-reactions_uniso_lum_alpha2}
\end{figure}

Finally, we compare the previous results, obtained assuming an
isotropic $\nu$ emission (Eq.~(\ref{eq: neutrino fluxes}) with $\alpha=0$), with the ones 
obtained in the anisotropic case (Eq.~(\ref{eq: neutrino fluxes}) with $\alpha=2$). 
While the entropy and the expansion velocity show only minor variations,
more interesting differences are visible in the electron fraction distributions, presented in
Figure~\ref{fig:ye-weak-reactions_uniso_lum_alpha2}. 
Since most of the ejection happens inside a solid angle delimited by a polar angle $\theta = \pi/2 \pm \pi/6$
about the equatorial plane, the assumption of an anisotropic neutrino emission decreases the neutrino fluxes
experienced by the escaping matter and lower the effect of electron neutrino absorption. 
This results in systematically more neutron-rich ejecta, with electron abundances
usually decreased by $\sim$15\% and average values located between $\langle \Ye \rangle \approx 0.16$ 
(low luminosity) and $\langle \Ye \rangle \approx 0.21$ (high luminosity).
Our results suggest also that, in the case of ejecta emitted closer to the polar axis, a high degree of
anisotropy in the neutrino emission could result in a large increase of the electron fraction
inside the polar region.

\subsection{Nucleosynthesis yields}

\begin{figure}[!htb]
 \centering
 \includegraphics[width=0.95\linewidth]{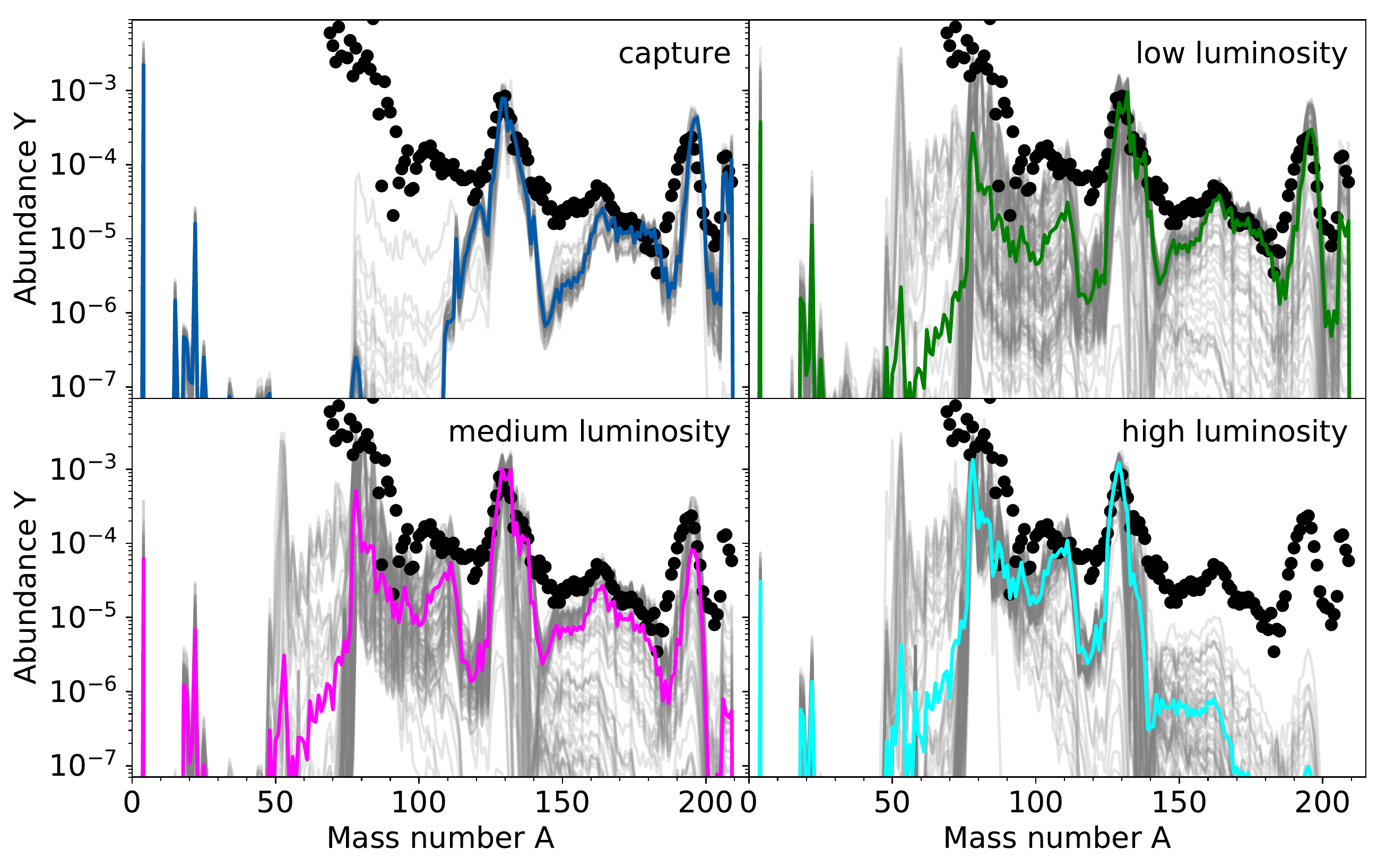}
 \caption{Nucleosynthesis yields when including weak reactions. We find a robust r-process for the case including only electron and positron captures. 
 However, the abundances for heavy nuclei with ($A \gtrsim 130$) decreases and the ones for lower mass nuclei ($A \lesssim 130$) increases 
 if higher (anti)neutrino luminosities are assumed. 
 Color scheme and luminosity cases are the same as in Figure~\ref{fig:ye-weak-reactions_iso_lum_alpha0}. 
 Gray lines show the abundance pattern of individual tracers. Solar r-process abundances are shown as red dots.}
 \label{fig:kepler_iso_lum_alpha0}
\end{figure}

Having post-processed the ejected tracers for obtaining an updated electron fraction evolution, 
we use the outcome as an input for subsequent nucleosynthesis calculations. The nucleosynthesis 
yields are shown in Figure~\ref{fig:kepler_iso_lum_alpha0}, alongside with the solar r-process abundances \citeaffixed{Lodders:2003}{dots, }.
We first consider the isotropic luminosity case,
$\alpha=0$ in Eq.~(\ref{eq: neutrino fluxes}). Gray lines represent the abundance patterns of individual tracers, 
while colored lines the abundances summed over all tracers.
For a straightforward comparison, we apply the same colors as for the electron fraction distributions 
in Figure~\ref{fig:ye-weak-reactions_iso_lum_alpha0}.
In the case with only electron and positron captures, we find a robust
r-process nucleosynthesis from second to third peak due to the
extremely neutron-rich conditions. Moreover, the abundances of
individual tracers are close to the average one, with a small spread reflecting the narrow distribution
in electron fraction. In contrast, the three remaining cases including neutrino captures show a rather strong dependence on the neutrino 
irradiation flux, and the larger spreads in the electron fraction distribution translate in a larger variety behaviors of the single tracers with
respect to the average one.
The component of the ejecta with relatively high electron fraction forms r-process nuclei up to the second peak.
When these ejecta are complemented by a neutron-rich component, the mass-integrated nucleosynthesis almost ranges from the first to the third 
r-process peak. The relative importance of the light to the heavy r-process nuclei depends on the increase in electron fraction and, in turn, on
the intensity of the neutrino luminosities. The first peak is somewhat underproduced in the case with low luminosities. 
Increasing the (anti)neutrino luminosities has two effects. First, the abundances of nuclei up to the second r-process peak are enhanced. 
Second, the abundances of heavy nuclei with $A \gtrsim 130$ decrease by up to more than an order of magnitude 
with respect to the solar abundances. In the high luminosity case, the increase of the 
electron fraction peak above $0.23$, combined with a rather low matter entropy, prevents a significant production of r-process nuclei above the second peak.

\begin{figure}[!htb]
 \centering
 \includegraphics[width=0.95\linewidth]{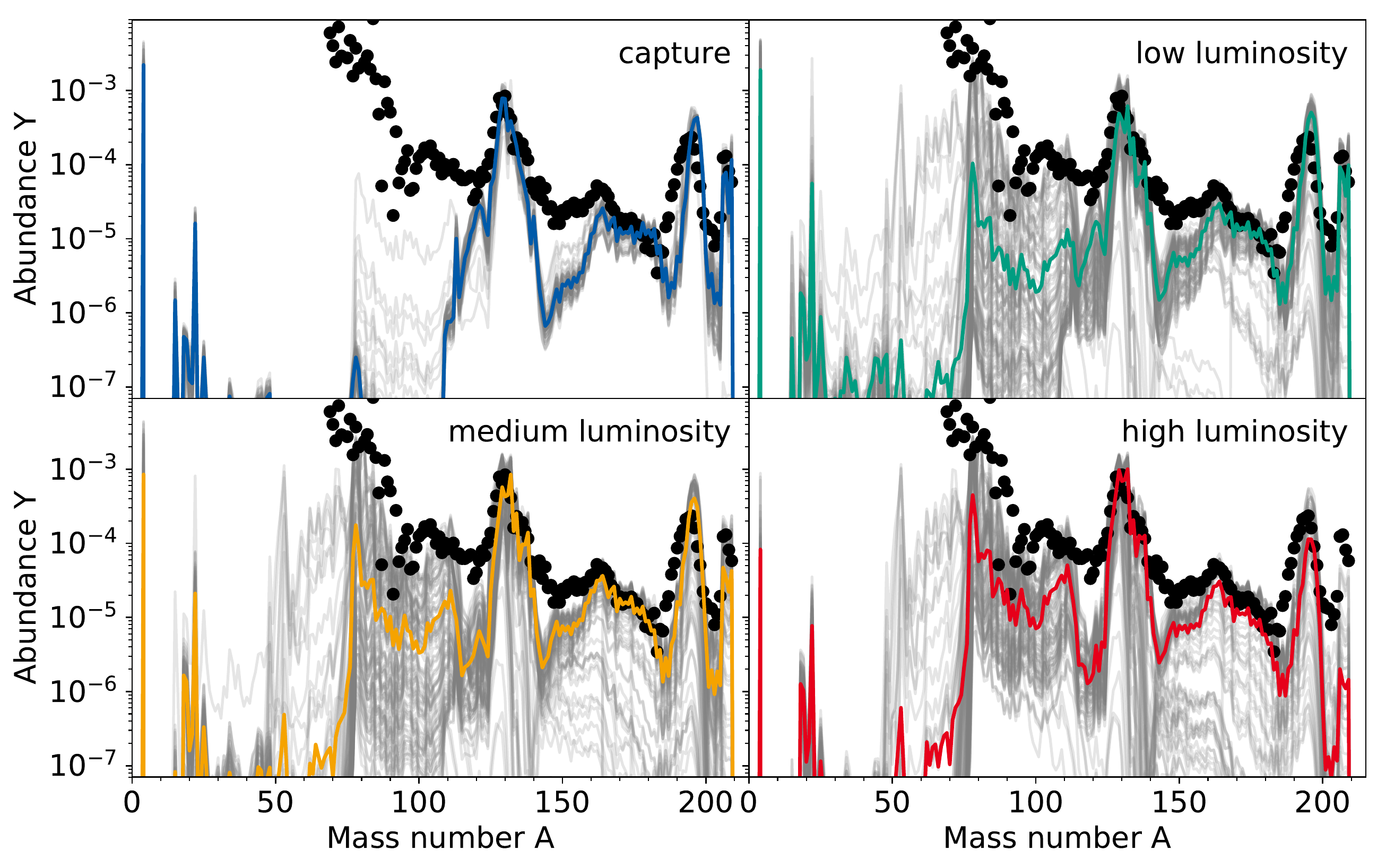}
 \caption{Nucleosynthesis yields when including weak reactions and angle dependent luminosities. Color scheme and luminosity cases are the same as in Figure~\ref{fig:ye-weak-reactions_uniso_lum_alpha2}.}
 \label{fig:kepler_uniso_lum_alpha2}
\end{figure}

The angular dependence of the (anti)neutrino luminosities also affects the r-process nucleosynthesis in the dynamic ejecta. 
In Figure~\ref{fig:kepler_uniso_lum_alpha2}, we present the final abundance yields obtained in the case of anisotropic neutrino emission,
$\alpha=2$ in Eq.~(\ref{eq: neutrino fluxes}), again in comparison with the solar r-process abundances \citeaffixed{Lodders:2003}{dots, }. 
Since the dynamic ejecta are mainly ejected close to the equatorial plane, 
the shadow effect of the forming disk reduces the neutrino fluxes
that irradiate the expanding matter. Similar to the isotopic cases, a weak component that comprises nuclei with $A \lesssim 130$ 
is coproduced. As a consequence, this reduces the abundances in the region beyond the second r-process peak, but not as much as 
in the cases exhibiting isotropic luminosity. On the contrary, we find a rather robust r-process pattern, which underproduces 
the rare-earth peak and the third peak by up to a factor of $\sim 2$ for the highest assumed (anti)neutrino luminosities. 
All yields are complemented by lighter heavy elements between the first and the second peak.

\subsection{Discussion and comparison}

Our work extends on the results of previous studies (e.g., \citeasnoun{Roberts.etal:2017} and, in particular, \citeasnoun{Goriely.etal:2015}). 
\citeasnoun{Goriely.etal:2015} report a similar initial\footnote{For beta-equilibrium conditions at 
$\rho_{\rm eq} = 10^{12}$g\,cm${^{-3}}$.} electron fraction distribution as we do ($0.0 \lesssim \Ye \lesssim 0.2$). 
However, they find a peaked instead of a bimodal distribution, and their distribution extends to $\Ye = 0.3$. Particularly when solely 
considering electron and positron captures in the subsequent evolution of $\Ye$, they find a
very broad distribution, and this is because of to the occurrence of a shock at larger densities and much larger temperatures 
($T \gtrsim 100\,{\rm GK}$). In their cases including (isotropic) neutrino irradiation, the relatively large (neutrinoless) weak equilibrium $\Ye$ occurring at weak 
freeze-out is further increased by $\nue$ and positron absorptions. Thus, when including also neutrino captures on nucleons, 
they report on higher electron fractions than we find in our results, even up to $\Ye = 0.5$ (though this is likely also due to the different 
treatment of $\lambda^k_{\nu}$). 
Their corner case for decreased temperatures ($T/3$) is comparable to our results, since the temperatures are similarly high here. 
The lower shock temperature generally implies a milder decrease of $\Ye$ during the shock, and therefore more emphasized 
effects of $\nue$ and $\anue$ captures, leading to an overall higher $\langle \Ye \rangle$ at late times. When only taking into 
account electron and positron captures, they find a distribution that is very similar to our corresponding case.
Overall, our results confirm their conclusion that the r-process nucleosynthesis shows a robust abundance pattern, 
unless the neutrino flux is extraordinarily high (see our case ``high luminosity''). Nevertheless, the weak r-process component 
with mass numbers $A \lesssim 130$ is affected by the degree of weak interactions, i.e., the larger the neutrino flux the more 
the ejecta yield nuclei below the second r-process peak. Under the assumptions made here, there are always admixtures of these 
two distinct components and there is a trade-off to produce either of the two. Therefore, we need at least a second kind of 
ejecta to explain the full r-process pattern from the first to the third r-process peak \citeaffixed{Hansen.etal:2014}{see, e.g.,}.

For the merger of a black hole and a neutron star, \citeasnoun{Roberts.etal:2017} investigate the impact of neutrinos on the nucleosynthesis.
It is important to note that such a system lacks an interaction region, hence the ejecta originate from the tidal tails. 
This component is extremely neutron-rich and cold enough for electron and positron captures to be neglected. 
Due to the rapid outflow timescales of the ejecta, \citeasnoun{Roberts.etal:2017} find that the electron fraction distribution 
is not shifted significantly. Even for their highest (isotropic) luminosity case 
($L_{\nue} = 2.5 \cdot 10^{53}\,\mathrm{erg/s}$, $L_{\anue} = 1.5 L_{\nue}$), the electron fraction obeys 
$\Ye \lesssim 0.25$. Similar to our results, the heavy elements from the second to the third r-process peak are 
produced robustly, with subtle enhancements for the abundances of nuclei with $A \lesssim 130$ due to neutrino irradiation.

Furthermore, GR radiation-hydrodynamics simulations also found that $\Ye$ in the dynamic ejecta of BNS mergers 
can be significantly enhanced by weak reactions \citeaffixed{Sekiguchi.etal:2015,Foucart.etal:2015,Radice.etal:2016}{e.g.,}. 
However, \citeasnoun{Sekiguchi.etal:2015} reported that the inclusion of neutrino 
absorption in optically thin conditions has a minor impact on the average electron fraction of the ejecta 
in their models. These could be explained by the occurrence of the shock at lower electron degeneracy conditions (i.e., at lower densities and/or higher
temperatures), favoring positron captures on neutrons, rather than electron captures on protons (see, for example, Figure~\ref{fig:critical-density-temperature}).

\section{Conclusion}
\label{sect:conclusion}

In this paper, we have explored the impact of weak reactions on the distributions of the electron fraction and of the entropy in the ejecta
obtained from an equal mass neutron star binary merger simulated in full GR, with a finite temperature, microphysical EOS.
We have focused on the shock-heated ejecta that originates from the disk.
We have used a parameterized post-processing treatment that allowed us to explore consistently the impact of individual reactions on
$\Ye$ and on the entropy.
It also permitted to disentangle the role of some of the most relevant
aspects influencing the electron fraction evolution, including the impact of shock heating, the dependence on the intensity of the integrated neutrino
luminosities, and on the degree of anisotropy in the neutrino emission. For each model, we have computed detailed nucleosynthesis yields, relating
the impact of weak reactions to the properties of the synthesized nuclear abundances.

These are our three major findings:
\begin{itemize}
 \item[1.)] The inclusion of neutrino absorption on free nucleons, in addition to neutrino emission from electron and positron absorptions,
  changes significantly the properties of the ejecta.
  Even if electron antineutrino luminosity initially dominates over electron neutrino luminosity, the neutron abundance and the
  larger reaction $Q$-value favor electron neutrino absorption on neutrons for all the tested luminosities, increasing always the electron fraction.
  The larger the neutrino luminosities are, the larger the increase in $\Ye$ distribution is.
  Moreover, the increase in matter entropy due to the absorption of neutrinos roughly compensate the decrease due to neutrino emission.
 \item[2.)] The occurrence of a shock in the ejection process does not necessarily lead to an increase of the electron fraction in neutron-rich matter.
  If the shock occurs at densities $ \gtrsim 10^{11} \, {\rm g \, cm^{-3}}$ and temperatures $\lesssim 8 \, {\rm MeV}$,
  electron degeneracy favors electron captures on protons rather than positron captures on neutrons, for an initial $\Ye > 0.10$.
  For the examined trajectories, when the peak temperature in the shock exceeds $\sim 5 \, {\rm MeV}$ and $\Ye > 0.1$, electron
  captures are fast enough to reach weak equilibrium around $\Ye \approx 0.10$. This has the remarkable consequence that the subsequent
  evolution becomes independent from the thermodynamical history before the shock.
  On the other hand, if $\Ye < 0.1$ before the shock occurs, positron absorption is not fast enough to ensure equilibrium.
  In our tracer sample, the latter condition is verified only when neutrino absorption is neglected. Otherwise, neutrino absorption increases
  always $\Ye$ above 0.1 before the shock occurrence.
 \item[3.)] Neutrino absorption is proportional to the local neutrino flux. Thus, in addition to the total luminosity and energy spectrum,
  the angular dependence of the neutrino luminosity, in combination with the spatial distribution of the ejecta, has a relevant
  impact. Since the shock heated ejecta that we have analyzed in our work expand close to the equatorial plane, a significant
  degree of anisotropy in the neutrino emission can lead to appreciable differences in the $\Ye$ distributions, compared with the isotropic case.
\end{itemize}

Our work confirms previous findings that weak reactions are crucial to set the properties of the ejecta in binary compact mergers, even for 
the dynamic ejecta \cite{Wanajo.etal:2014,Sekiguchi.etal:2015,Goriely.etal:2015,Sekiguchi.etal:2016,Radice.etal:2016,Roberts.etal:2017}. 
Thus, future studies that aim at exploring the properties of the ejecta and address the problem of the related nucleosynthesis 
will require a careful inclusion of neutrino physics, both in terms of the relevant neutrino reactions and of the characteristic emission properties.
The detailed results we have obtained are intrinsically related with the specific properties of the tracer particles we have used in our
analysis. Some of our findings might not apply to shock-heated ejecta that significantly differ from ours. 
However, we have shown that our approach is useful to investigate the origin of the increase in $\Ye$ in decompressed neutron-rich matter
from binary compact mergers, as well as the relevance of single reactions.
Moreover, it is well suited to analyze the results of detailed radiation-hydrodynamical simulations with a controlled
and inexpensive approach, in particular to explore in details the different thermodynamical 
conditions experienced by fluid elements during a binary merger.
A larger and more detailed set of models is required to extensively explore the different conditions experienced by matter 
and radiation during compact mergers.

\ack
The authors thank L. Roberts, and D. Radice for useful discussions.
A. A., D.M. and A. P. acknowledge support from the Helmholtz-University Investigator grant No. VH-NG-825, from the BMBF under grant No.
05P15RDFN1, and from the European Research Council Grant No. 677912 EUROPIUM.
A. P. and D. M. thank the GSI Helmholtzzentrum f\"ur Schwerionenforschung GmbH for the usage of computational resources.
The simulations with the \texttt{FISH+ASL} code were supported by a grant 
from the Swiss National Supercomputing Centre (CSCS) under project ID 667.

\appendix

\section{Weak magnetism and recoil corrections. Approximated captures rate expressions.}
\label{ap:weak_magn_corr_and_approx_rates}
In Eqs.~(\ref{eq:rate-electron-capture})$-$(\ref{eq:rate-anue-capture}), we employ the weak magnetism and recoil corrections
provided by \citeasnoun{Horowitz:2002} for charged current reactions on free nucleons:
\begin{eqnarray}
\mathcal{R}_{\nu}(E) & = \frac{1}{c_v^2+ 3 c_a^2} \frac{1}{\left( 1 + 2x \right)^3}
\left[ c_v^2 \left( 1 + 4x + \frac{16}{3}x^2 \right) + 3 c_a^2 \left( 1 + \frac{4}{3}x \right)^2 \right. \nonumber \\
 & \left. \pm 4 \left( c_v + F_2 \right) c_a x \left( 1 + \frac{4}{3}x \right)
+ \frac{8}{3} c_v F_2 x^2 + \frac{5}{3} x^2 \left( 1 + \frac{2}{5} x \right) F_2^2 \right] \, .
\label{eq:weak_mag_expression}
\end{eqnarray}
In the above expression, $x=E/(M_{\rm b}c^2)$, $M_{\rm b}$ is the baryon mass, $c_v=1$, $c_a = g_a \approx 1.26$, and $F_2 \approx 3.706$.
The upper sign refers to $\nue$, the lower to $\anue$.

In our calculations, we did not consider any approximations to the rates, while in the following,
starting from Eqs.~(\ref{eq:rate-electron-capture})$-$(\ref{eq:rate-anue-capture}),
we derive approximated expressions in the form:
\begin{equation}
 \lambda^0_{x} = c ~ n_x ~ \langle \sigma_x \rangle \, ,
 \label{eq: lambda rates} \\
\end{equation}
where $n_x$ is the target density and $\langle \sigma_x \rangle$ an average cross section.
This derivation is useful to provide simpler expressions for the rates and to compare with others used in the literature. 
We neglect both the electron rest mass correction ($\mathcal{M} \rightarrow 1$ and $\Delta+m_\e \rightarrow \Delta$
in the non-vanishing lower limits of integration) 
and the Pauli blocking factors involving electrons and positrons in the final state.
Moreover, since for typical neutrino energies $x \ll 1$, we expand Eq.~(\ref{eq:weak_mag_expression}) in powers of $x$:
\begin{eqnarray}
\mathcal{R}_{\nu}(E) & \approx 1  -
2 x \frac{\left( 5 c_a^2 \mp 2 c_a \left( c_v + F_2 \right) + c_v^2 \right)}{\left( c_v^2+ 3 c_a^2 \right)} + \nonumber \\
& \frac{1}{3}x^2 
\frac{\left( 88 c_a^2 \mp 56 c_a \left( c_v+ F_2 \right) + 5 F_2^2 + 8 c_v F_2 + 16 c_v^2  \right)}{ \left( c_v^2+ 3 c_a^2 \right)} + \mathcal{O}(x^3) \, .
\label{eq:weak_mag_expansion}
\end{eqnarray}

In the case of (anti)neutrino capture rates, $\lambda^0_{\nue}$ and $\lambda^0_{\anue}$, we assume the free streaming 
radiation to propagate mainly radially, 
and its spectrum to be described by a Fermi-Dirac distribution with vanishing degeneracy parameter 
and mean energy $\langle E_\nu \rangle = \kB T_{\nu} F_3(0)/F_2(0)$.
For the corresponding capture rates, we obtain
\begin{eqnarray}
 \lambda^0_{\nue} = c ~ n_{\nue} ~ \langle \sigma_{\nue} \rangle \, , \\
 \lambda^0_{\anue} = c ~ \tilde{n}_{\anue} ~ \langle \sigma_{\anue} \rangle \, . 
 \label{eq:lambda-nue-anue-rates}
\end{eqnarray}
In the above expressions, $n_{\nue}$ is the electron neutrino particle density, which can be expressed in terms of the local radial flux, 
Eq.~(\ref{eq: neutrino fluxes}), while $\tilde{n}_{\anue}$ is a modified expression of the electron antineutrino 
particle density, which takes into account the non-zero lower integration limit:
\begin{eqnarray}
n_{\nue} = \frac{\F_{\nue}}{c} \, , \\
\tilde{n}_{\anue} = \frac{\F_{\anue}}{c} \frac{F_2(-\Delta/ \kB T_{\anue})}{F_2(0)} \, .
\end{eqnarray}
$F_k(\eta)$ is the Fermi integral of order $k$ and argument $\eta$.

The average cross sections $\langle \sigma_\nu \rangle$ are then computed by inserting Eq.~(\ref{eq:weak_mag_expansion}) up
to the first order in $x$ inside Eqs.~(\ref{eq:rate-nue-capture}) and (\ref{eq:rate-anue-capture}), and then averaging over the neutrino distribution 
functions:
\begin{eqnarray}
\langle \sigma_{\nu} \rangle & \approx \frac{\sigma_0}{\left( m_\e c^2 \right)^2} \left< \epsilon_{\nu}^2 \right>  
 \left\{ \left[ 1 + 2 \frac{\Delta \langle \epsilon_{\nu} \rangle}{\langle \epsilon_{\nu}^2 \rangle} + \frac{\Delta^2}{ \langle \epsilon_\nu^2 \rangle} \right] + \right.  \nonumber \\
 &  - \left. 
 \left( \gamma \mp \delta \right) \left[ \frac{\langle \epsilon_{\nu}^3 \rangle}{\langle \epsilon_{\nu}^2 \rangle M_b c^2}
 + 2 \frac{\Delta}{M_b c^2} + \frac{\Delta^2 \langle \epsilon_{\nu} \rangle}{\langle \epsilon_{\nu}^2 \rangle M_b c^2} \right]
 \right\} \, , \label{eq:ye-nu-cs} 
 \end{eqnarray} 
In Eq.~(\ref{eq:ye-nu-cs}), the upper sign is for $\nue$ while the lower sign for $\anue$, 
$\gamma=2\left( c_v^2 + 5 c_a^2 \right)/\left( c_v^2 + 3 c_a^2 \right)$, and
$\delta=4 c_a^2 \left( c_v + F_2 \right)/\left( c_v^2 + 3 c_a^2 \right)$ .
We compute the neutrino and antineutrino energy moments via
\begin{eqnarray} \label{eq:ye-enu-moments}
 \left< \epsilon_{\nue}^n \right> = (\kB T_{\nue})^n ~\frac{F_{n+2}(0)}{F_2(0)} \, , \\
 \left< \epsilon_{\anue}^n \right> = (\kB T_{\anue})^n ~\frac{F_{n+2}(-\Delta/(\kB T_{\anue}))}{F_2(-\Delta/(\kB T_{\anue})} \, ,
\end{eqnarray}
where we took again into account the lower integration limit by modifying the neutrino degeneracy parameter.

In the case of electron and positron captures on nucleons, the rates are expresses as
\begin{eqnarray}
 \lambda^0_{\e^-} = c ~ \tilde{n}_{\e^-} ~ \langle \sigma_{\e^{-}} \rangle \, , \label{eq:lambda-el-rates} \\
 \lambda^0_{\e^+} = c ~ n_{\e^+} ~ \langle \sigma_{\e^{+}} \rangle \, . 
 \label{eq:lambda-pos-rates}
\end{eqnarray}
In the previous expressions, $n_{\e^+}$ is the positron particle density while $\tilde{n}_{\e^-}$ is a modified 
version of the electron particle density, 
\begin{eqnarray}
 n_{\e^-} &= \frac{8\pi}{(2 \pi \hbar c)^3} \left(\kB T \right)^3  F_2 \left( \frac{\mu_\e - \Delta}{\kB T} \right)  \,, \label{eq:ye-el-density} \\
 n_{\e^+} &= \frac{8\pi}{(2 \pi \hbar c)^3} \left(\kB T \right)^3  F_2 \left( - \frac{{\mu_\e}}{\kB T} \right) \label{eq:ye-pos-density} \,,
\end{eqnarray}
where $\mu_{\e}$ is the relativistic electron chemical potential, and we used the fact that for high enough temperatures 
$\mu_{\e^+} = - \mu_{\e}$.
The average capture cross-sections for electrons and positrons, $\langle \sigma_{\e^{-}} \rangle$ and $\langle \sigma_{\e^{+}} \rangle$, 
are given by:
\begin{eqnarray}
\langle \sigma_{\e} \rangle & \approx \frac{\sigma_0}{ 2 \left( m_\e c^2 \right)^2} \left< \epsilon_{\e}^2 \right>  
 \left\{ \left[ 1 + 2 \frac{\Delta \langle \epsilon_{\e} \rangle}{\langle \epsilon_{\e}^2 \rangle} + \frac{\Delta^2}{ \langle \epsilon_\e^2 \rangle} \right] + \right.  \nonumber \\
 &  - \left.
 \left( \gamma \mp \delta \right) \left[ \frac{\langle \epsilon_{\e}^3 \rangle}{\langle \epsilon_{\e}^2 \rangle M_{\rm b} c^2}
 + 2 \frac{\Delta}{M_{\rm b} c^2} + \frac{\Delta^2 \langle \epsilon_{\e} \rangle}{\langle \epsilon_{\e}^2 \rangle M_{\rm b} c^2} \right]
 \right\} \, .
\label{eq:ye-el-cs}
\end{eqnarray}
where the upper sign refer to $\e^-$ and the lower sign to $\e^+$.
The energy moments are computing in terms of the Fermi integrals and also in this case for $\e^-$ the non vanishing lower integration limit 
is included as a shift in the chemical potential,  
\begin{eqnarray}
 \left< \epsilon_{\e^-}^n \right> &= (\kB T)^n~\frac{F_{n+2}(\left( \mu_\e - \Delta \right)/(\kB T))}{F_{2}( \left( \mu_\e - \Delta \right)/(\kB T))} \,, \label{eq:ye-el-moments} \\
 \left< \epsilon_{\e^+}^n \right> &= (\kB T)^n~\frac{F_{n+2}(-\mu_\e/(\kB T))}{F_{2}(-\mu_\e/(\kB T))} \,. \label{eq:ye-pos-moments}
\end{eqnarray}
We notice that the expressions we have derived for $\langle \sigma_{\nu} \rangle$ are similar, but different from the expressions reported in
\citeasnoun{Horowitz.Li:1999}. 

\begin{figure}[!htb]
 \centering
 \includegraphics[width=\linewidth]{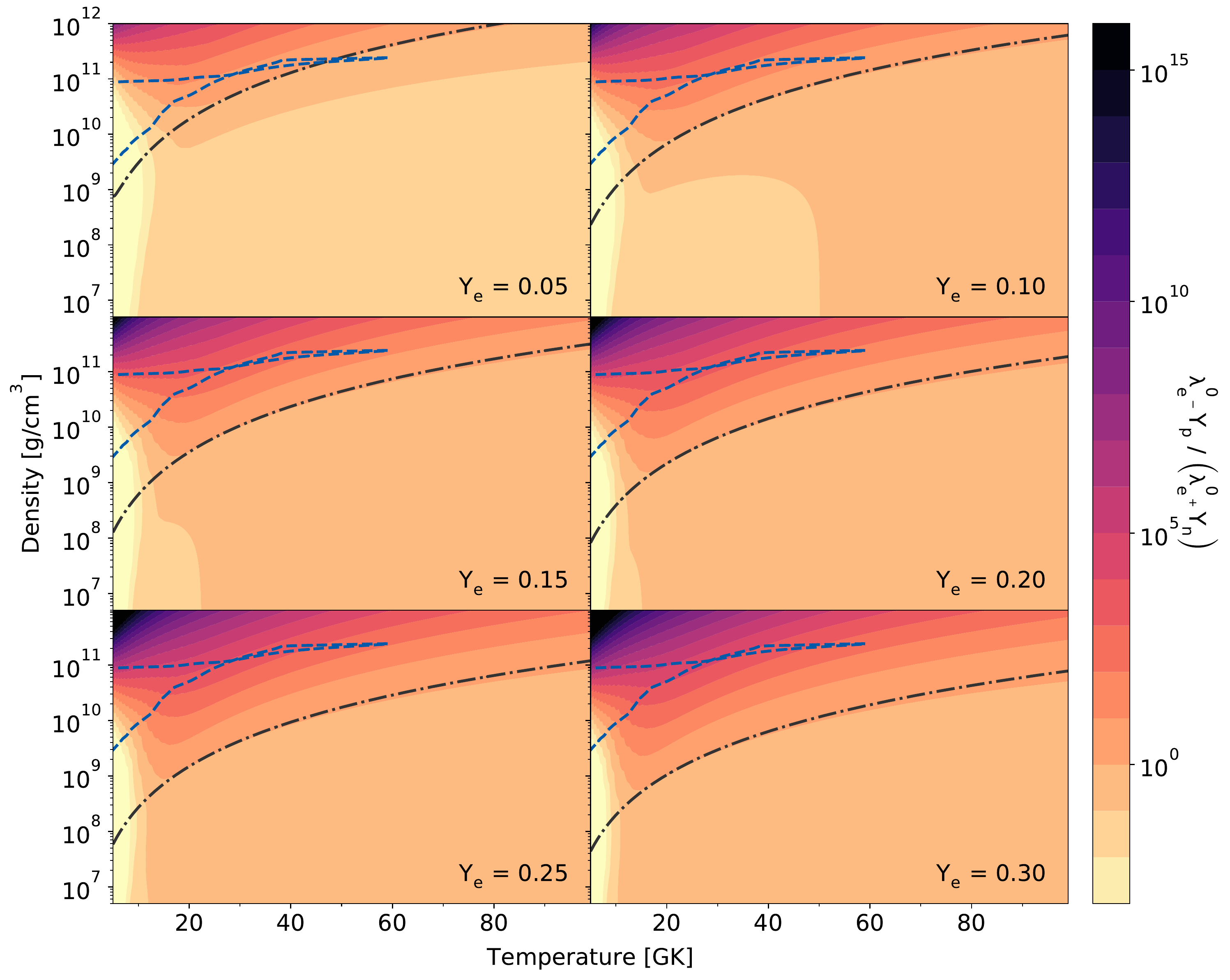}
 \caption[Ratio of the electron and positron capture rates.]{Ratio of the electron and positron capture rates
 in the density-temperature plane for varying electron fractions. We plot the trajectory from Figure~\ref{fig:critical-density-temperature} 
 with a dashed blue line.  The dotted-dashed black line marks the solution of Eqs.~(\ref{eq:ratio-elpos-captures}) 
 for a composition made up of only nucleons.}
 \label{fig:ratio-elpos-captures}
\end{figure}

In Figure~\ref{fig:critical-density-temperature}, we have presented curves of equal electron and positron capture rates in the matter density-temperature
plane. In Figure~\ref{fig:ratio-elpos-captures}, we plot for completeness the ratio between the two rates over the full plane.

\section{Model for atmospheric drag} \label{sec:drag}
In the following, we construct a model for the atmospheric drag applicable 
to the ejecta considered in this work. Those are arranged in two expanding 
thin rings above and below the orbital plane. 
At the relevant distances from the remnant, Newtonian physics is sufficient.
We assume that each fluid element 
of the ring would follow an unbound orbit without the drag force provided by the 
artificial atmosphere. To compute the latter, we assume 
that artificial atmosphere swept up by the expanding ring becomes part of it, while
the linear and angular momentum of infinitesimal ring sections remains 
unchanged by this merging. We regard this as a reasonable 
approximation for the behavior of the finite volume numerical scheme unless 
the ring density becomes comparable to the atmosphere density.
Given the atmosphere density $\rho_a$ and an effective projected area $A_r$ of 
the ring, the increase in the mass $m$ of the ring is $\dot{m} = A_r \rho_a v_s$, where $v_s$
is the velocity of the ring surface. We further approximate $v_s \approx \dot{r}$,
where $r^2 = R_r^2 + z_r^2$, where $R_r$ and $z_r$ are the ring radius and 
the $z$-offset of the ring.
To get the projected area, we take the increase in thickness during the expansion into 
account by making the assumption $A_r \approx  4\pi a r^2$, where 
$a$ is a constant denoting the effective fraction of the solid angle covered by ejecta. 
This holds exactly if the expansion can be described 
as uniform scaling.

\begin{figure}
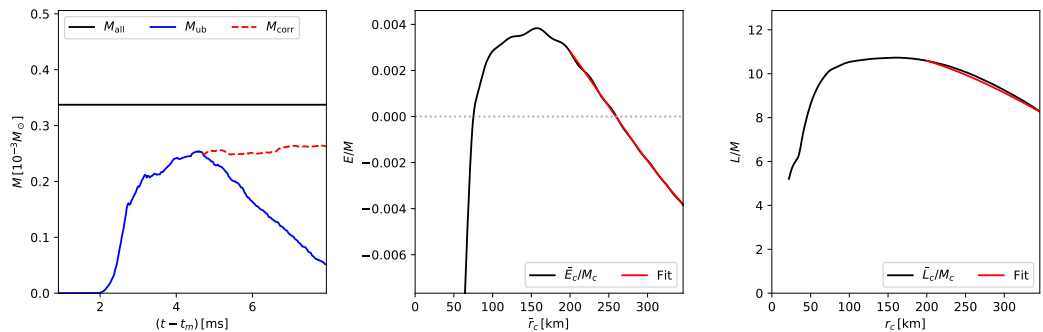

  \begin{center}
    \includegraphics[width=0.9\columnwidth]{{{drag_fit}}} 
    \caption{Influence of artificial atmosphere on ejecta. Left: total mass of tracers (black), tracer 
    mass unbound at a given time according to geodesic criterion (blue), after correcting for the energy 
    loss according to drag model fit (red). Middle: kinetic plus potential energy of 
    tracers divided by total tracer mass (black), and the drag model fit (red). Right: total angular
    momentum of tracers divided by total tracer mass (black), and the drag model fit (red). }
    \label{fig:drag}
  \end{center}
\end{figure}

Combining all the above assumptions, we obtain 
$\dot{m} = 4\pi a  r^2 \rho_a \dot{r}$.
Using momentum and angular momentum conservation and a central gravitating 
mass $M$, a short computation yields
\begin{eqnarray}
\frac{{\rm d}}{{\rm d}t} \left(\frac{E}{m} \right) &= \frac{{\rm d}}{{\rm d}t} \left(\frac{E_k}{m} - \frac{M}{r} \right) = - 2 \frac{\dot{m}}{m} \frac{E_k}{m}, \qquad
l^z = l^z_0 \frac{m_0}{m} \, ,
\end{eqnarray}
where $E_k$ is the kinetic energy of the ring and $l_z$ the specific angular momentum.
Rewriting $m(t)$ as $m(r)$, we find
\begin{equation}
\frac{{\rm d}m}{{\rm d}r}  = 4\pi a  r^2 \rho_a, \qquad
\frac{{\rm d}}{{\rm d}r} \left(m E \right) = -8\pi a \rho_a M m r \, ,
\end{equation}
Integration yields the result
\begin{eqnarray}
m &= m_0 \left( 1 + \frac{k_1}{3} \left(r^3 -r_0^3 \right)\right) \\
\frac{E}{m} &= \frac{m_0^2}{m^2} \left[ 
                \frac{E_0}{m_0} + M k_1 \left(\left(\frac{1}{3} k_1 r_0^3 - 1\right)
\left(r^2 - r_0^2 \right) - \frac{2}{15} k_1 \left(r^5 - r_0^5 \right)\right) \right] \, ,
\end{eqnarray} 
where the subscripts ${}_0$ denote initial values and  $k_1 = 4\pi a \rho_a/m_0$.
A fit of $k_1$ to the extracted trajectories is shown in the middle panel of 
Figure~\ref{fig:drag} (we picked a starting time well after the ring became unbound). 
Next, we computed the angular momentum using the fit parameter obtained 
above. The result is shown in the right panel, and fits the data sufficiently well.
Our main requirement for the fit is however that it can account for the loss of
unbound matter. To test this, we assume that the loss of (kinetic plus potential) 
specific energy for each ejecta fluid element is a function of radius given by the 
fit. We can then adjust the energy threshold for the geodesic bound matter criterion
to correct for this energy loss. The result is shown in the left panel. 
After the correction, the sharp decrease is removed. Instead, we find
a slight increase, which is likely an over-correction due to the various 
approximations. To avoid confusion, we note that the mass shown in the panel
is not $m$ in the equations above, but a sum over the (constant) tracer 
masses of the tracers which are unbound at a given time, i.e. it should remain constant
if no tracer becomes bound/unbound anymore.
In conclusion, our model supports the assumption that the slow-down of our ejecta
is indeed just a numerical artifact, as assumed in \cite{Kastaun.etal:2016b} to compute the ejecta mass.
The model might also be useful to plan numerical simulations since it allows to
predict the slowing of the ejecta for given (constant) atmosphere density.

\bibliographystyle{jphysicsB}

\end{document}